\documentclass[final,3p,times,fleqn]{elsarticle}
\usepackage{mathtools,booktabs}
\usepackage{bm}
\usepackage{autobreak}
\usepackage{amsmath}
\usepackage{amsthm}
\usepackage{gensymb}
\usepackage{subfigure}
\usepackage{epstopdf}
\usepackage{color}
\usepackage{hyperref}
\newtheorem{remark}{Remark}
\biboptions{sort&compress}
\begin{document}
	\begin{frontmatter}
		\title{A diffuse-interface lattice Boltzmann method for the dendritic growth with thermosolutal convection}
		\author[a]{Chengjie Zhan}
		\author[a,b,c]{Zhenhua Chai \corref{cor1}}
		\ead{hustczh@hust.edu.cn}
		\author[a,b,c]{Baochang Shi}
		\author[d]{Ping Jiang}
		\author[d]{Shaoning Geng}
		\author[e]{Dongke Sun}
		\address[a]{School of Mathematics and Statistics, Huazhong University of Science and Technology, Wuhan 430074, China}
		\address[b]{Institute of Interdisciplinary Research for Mathematics and Applied Science, Huazhong University of Science and Technology, Wuhan 430074, China}
		\address[c]{Hubei Key Laboratory of Engineering Modeling and Scientific Computing, Huazhong University of Science and Technology, Wuhan 430074, China}
		\address[d]{The State Key Laboratory of Digital Manufacturing Equipment and Technology, School of Mechanical Science and Engineering, Huazhong University of Science and Technology, Wuhan 430074, China}
		\address[e]{School of Mechanical Engineering, Southeast University, Nanjing 211189, China}
		\cortext[cor1]{Corresponding author.}
		\begin{abstract}
			In this work, we proposed a diffuse interface model for the dendritic growth with thermosolutal convection. In this model, the sharp boundary between the fluid and solid dendrite is replaced by a thin but nonzero thickness diffuse interface, which is described by the order parameter governed by the phase-field equation for the dendritic growth. The governing equations for solute and heat transfer are modified such that the previous special treatments for source term can be avoided. 
			To solve the model for the dendritic growth with thermosolutal convection, we also developed a diffuse-interface multi-relaxation-time lattice Boltzmann (LB) method.   
			In this method, the order parameter in the phase-field equation is combined into the force caused by the fluid-solid interaction, and the treatment on the complex fluid-solid interface can be avoided. In addition, four LB models are developed for the phase-field equation, concentration equation, temperature equation and the Navier-Stokes equations in a unified framework. Finally, to test the present diffuse-interface LB method, we performed some simulations of the dendritic growth, and found that the numerical results are in good agreements with some previous works.     
		\end{abstract}
	\end{frontmatter}
\section{Introduction}
Dendritic growth, as a complicated phase transition process coupling with melt flow, heat and solute transfer, is usually observed in both nature and engineering. To reveal the underlying mechanism of morphological evolution and to improve the properties of materials during the solidification process, the problem of dendritic growth has been investigated extensively from experimental and numerical perspectives \cite{Reinhart2014JOM,Casari2016AM,Stefan2018MMTA,Szep1985JPA,Harris1990JPA}. 

With the development of computer science and scientific computing, the numerical simulation has become a powerful and important tool in the study of the solidification processes, and a large number of mathematical models and numerical methods have been developed for the dendritic growth, for instance, the cellular automation method \cite{Zhu2007AM,Luo2013CMS}, enthalpy method \cite{Voller2008IJHMT,Bhattacharya2014JCP}, level-set method \cite{Tan2007JCP} and phase-field method \cite{Chen2002ARMR,Boettinger2002ARMR}.
The phase-field method with an order parameter introduced to distinguish different phases, has been widely used to simulate dendritic growth \cite{Asta2009AM,Steinbach2013JOM} for its thermodynamically self-consistence and needless of explicit interface-tracking. In the early works \cite{Beckermann1999JCP,Ramirez2004PRE,Yuan2010MSMSE}, however, the phase-field method is only adopted for the dendritic growth with pure diffusion, and the effects of thermal convection, solutal convection and melt flow are usually considered separately. In addition, the Navier-Stokes (NS) equations for flow field are solved by some traditional computational fluid dynamics approaches, which may also bring some difficulties in treating fluid-solid interaction and complex boundary conditions.

In the past three decades, the mesoscopic lattice Boltzmann (LB) method, as a popular kinetic-theory based numerical approach, has become an efficiently numerical tool in the simulation of complex fluid flows \cite{Higuera1989EPL,Benzi1992PR,Chen1998ARFM,Aidun2010ARFM} and nonlinear systems \cite{Shi2009PRE,Chai2013PRE}, such as the multiphase and multicomponent flows \cite{Liang2018PRE,Wang2019Capillarity}, phase transitions \cite{Sun2009AM,Zhao2019IJHMT} and fluid flows in porous media \cite{Chen1998ARFM,Succi2001}. Compared to the traditional computational fluid dynamics approaches, the LB method has some distinct features, including the simplicity in coding, easy implementation of complex boundary conditions, and fully parallel algorithms \cite{Chen1998ARFM}.
Considering the advantages of the phase-field and LB methods, some phase-field LB models for the dendritic growth have been developed \cite{Samanta2022CMS,Medvedev2005PRE,Sun2009AM,Rojas2015JCP,Sun2016IJHMT,Zhang2018CPC,Cartalade2016CMA,Sun2019IJHMT,Wang2021CMAME}. However, in the most of previous works on phase-field LB models \cite{Medvedev2005PRE,Sun2009AM,Rojas2015JCP,Sun2016IJHMT,Takaki2017JCG,Zhang2018CPC}, the LB method is only used to simulate melt flows as well as heat and solute transfer, and other techniques are adopted to solve the phase-field equation. To develop the efficient computational model in a unified LB framework, Cartalade et al. \cite{Cartalade2016CMA,Younsi2016JCP} proposed an anisotropic LB model for the dendritic growth with interfacial anisotropy, and the streaming is modified to achieve the relaxation time in the phase-field equation. Sun et al. \cite{Sun2019IJHMT,Sun2020AML} extended the anisotropic LB model to study thermal dendritic growth in the presence of melt flow, in which the multi-relaxation-time (MRT)-LB model is applied to simulate the melt flow, and the no-slip boundary condition is incorporated into the evolution equation through treating the diffusive liquid-solid interface as a kind of porous medium. Later, a hybrid method was applied to model the isothermal crystal growth in the solidification of binary alloys \cite{Wang2020CMS} where the LB and finite-volume schemes are used to solve phase-field and solute transfer equations, respectively. Recently, Wang et al. \cite{Wang2021CMAME} presented a phase-field MRT-LB model for the dendritic growth coupled with thermosolutal convection, and also developed a local computing scheme for the gradient of order parameter. Additionally, the multiple-time-scaling strategy was also used to improve the numerical stability when a wide range of physical parameters are considered in their simulations.

Although some LB models have been developed for the dendritic growth \cite{Cartalade2016CMA,Sun2019IJHMT,Wang2021CMAME}, there are some critical problems needed to be addressed. The first is that for the governing equations considered in the previous works \cite{Cartalade2016CMA,Sun2019IJHMT,Wang2020CMS,Wang2021CMAME}, the time derivative of order parameter and that of temperature or supersaturation are treated separately, and thus some other difference schemes should be used to discretize these derivatives. The second is how to treat the fluid-solid interaction accurately and efficiently in the framework of LB method. To solve these two problems, in this work we first introduced a new variable and modified the governing equations, then proposed a diffuse-interface MRT-LB method for dendritic growth where the fluid-solid interaction can be treated efficiently.

The remainder of this paper is organized as follows. In Section \ref{GoverEqs}, the governing equations for the thermosolutal dendritic growth with melt flow are introduced, and followed by the diffuse-interface MRT-LB method in Section \ref{LBMs}. In Section \ref{Numerical}, the numerical validations and discussion are presented, and finally, some conclusions are summarized in Section \ref{Conclusion}. 

\section{The diffuse-interface model for dendritic growth}\label{GoverEqs}
\begin{figure}
	\centering
	\subfigure{
		\begin{minipage}{0.49\linewidth}
			\centering
			\includegraphics[width=3.5in]{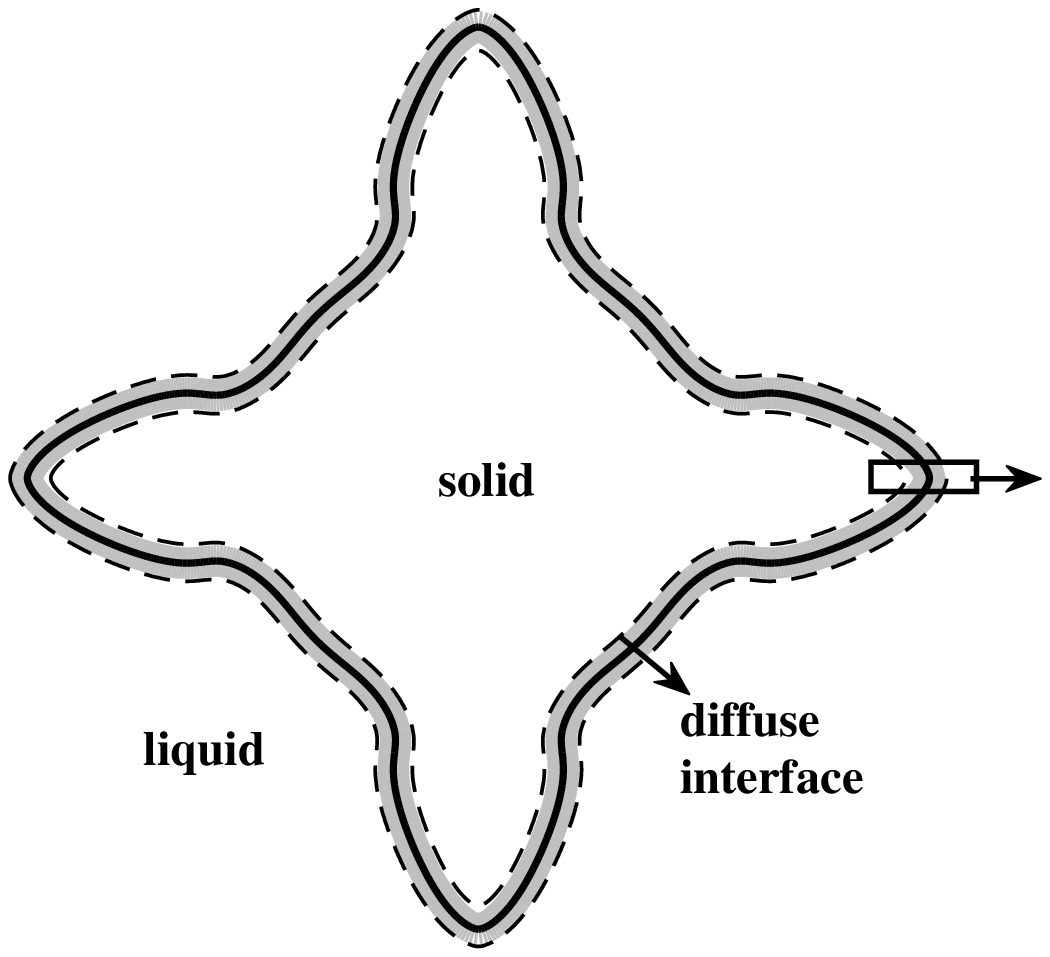}
	\end{minipage}}	
	\subfigure{
		\begin{minipage}{0.49\linewidth}
			\centering
			\includegraphics[width=2.5in]{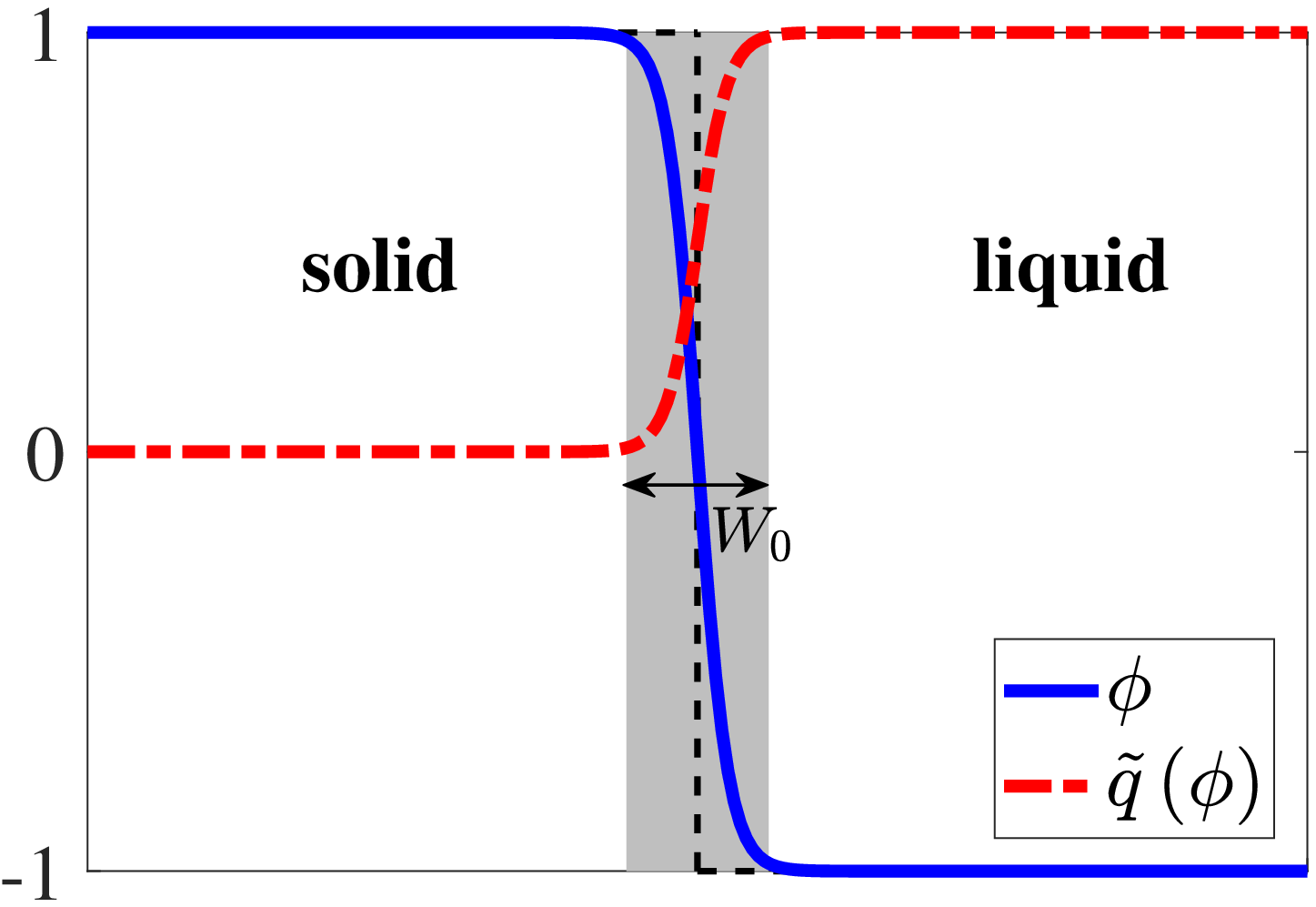}
	\end{minipage}}	
	\caption{The diffuse interface representation of the dendrite (left) and the profile of the phase-field variables (right).}
	\label{fig-diffuse}
\end{figure}

In the phase-field method, an order parameter $\phi$, smoothly changed from 1 in solid phase to -1 in liquid phase, can be applied to depict the fluid-solid interface ($\phi=0$), as shown in Fig. \ref{fig-diffuse}. Then the governing equation for phase field can be expressed as \cite{Karma2001PRL,Echebarria2004PRE}
\begin{equation}\label{GoverPhi}
	\tau_0a_s^2\left(\mathbf{n}\right)\frac{\partial\phi}{\partial t}=W_0^2\nabla\cdot\left[a_s^2\left(\mathbf{n}\right)\nabla\phi+\mathbf{N}\right]+Q_{\phi}.
\end{equation}
Here $\tau_0$ is a kinetic characteristic time, $W_0$ is the thickness of the interface. For a cubic system of crystal growth, the anisotropy function of the interfacial energy $a_s\left(\mathbf{n}\right)$ is given by \cite{Echebarria2004PRE}
\begin{equation}
	a_s\left(\mathbf{n}\right)=\left(1-3\varepsilon_s\right)\left[1+\frac{4\varepsilon_s}{1-3\varepsilon_s}\sum_{\alpha=1}^{d}n_{\alpha}^4\right],
\end{equation}
where $n_{\alpha}$ is the $\alpha$-component of the normal vector $\mathbf{n}=-\nabla\phi/|\nabla\phi|$ directing from solid to liquid, $d$ represents the dimensionality, and $\varepsilon_s$ is anisotropic strength. 
$\mathbf{N}=(N_{\alpha})$ is an anisotropic vector related to $a_s\left(\mathbf{n}\right)$, 
\begin{equation}
	N_{\alpha}=|\nabla\phi|^2a_s\left(\mathbf{n}\right)\frac{\partial a_s\left(\mathbf{n}\right)}{\partial\left(\partial_{\alpha}\phi\right)},\quad \alpha=1,2,\cdots,d.
\end{equation}

To extend the solute equation to the entire domain (solid, liquid and interface), the following dimensionless expressions of concentration $c$ and temperature $T$ are used \cite{Ramirez2004PRE,Cartalade2016CMA},
\begin{subequations}
	\begin{equation}
		U=\frac{\frac{2c/c_{\infty}}{1+k-\left(1-k\right)\phi}-1}{1-k},
	\end{equation}
	\begin{equation}
		\theta=\frac{T-T_m-mc_{\infty}}{L_h/C_p},
	\end{equation}
\end{subequations}
where $c_{\infty}$ is the initial composition of the melt, $k=c_s/c_l$ is the partition coefficient related to the compositions of solid and liquid in contact with each other at the interface, $T_m$ represents the melting temperature, $m$ donates the slope of the liquidus line in the phase diagram, $L_h$ is the latent heat and $C_p$ is the specific heat. With above dimensionless definitions, the coupling source term $Q_{\phi}$ in Eq. (\ref{GoverPhi}) can be defined by
\begin{equation}
	Q_{\phi}=\phi\left(1-\phi^2\right)-\lambda\left(\theta+Mc_{\infty}U\right)\left(1-\phi^2\right)^2,
\end{equation}
where $\lambda$ is the coupling coefficient, $M=-m\left(1-k\right)C_p/L_h$.

Subsequently, the governing equations for heat and solute transfer can be written as \cite{Karma2001PRL,Karma1998PRE}
\begin{subequations}\label{GoverPre}
	\begin{equation}
		\frac{1}{2}\left[1+k-\left(1-k\right)\phi\right]\frac{\partial U}{\partial t}=\nabla\cdot\left[D\tilde{q}\left(\phi\right)\nabla U-\mathbf{J}_{at}\right]+\frac{1}{2}\left[1+\left(1-k\right)U\right]\frac{\partial\phi}{\partial t},
	\end{equation}
	\begin{equation}
		\frac{\partial\theta}{\partial t}=\alpha\nabla^2\theta + \frac{\partial h\left(\phi\right)}{2\partial t},
	\end{equation}
\end{subequations}
where $\tilde{q}\left(\phi\right)=\left(1-\phi\right)/2$ is an interpolation function, $D$ is the solutal diffusivity in liquid, and is assumed to be zero in the solid. $\alpha$ is the thermal diffusivity, $h\left(\phi\right)=\phi$ or $15\left(\phi-2\phi^3/3+\phi^5/5\right)/8$ \cite{Karma1998PRE}. $\mathbf{J}_{at}$ denotes the phenomenological anti-trapping current term \cite{Karma2001PRL,Cartalade2016CMA}, and is determined by
\begin{equation}
	\mathbf{J}_{at}=-\frac{1}{2\sqrt{2}}W_0\left[1+\left(1-k\right)U\right]\frac{\partial\phi}{\partial t}\frac{\nabla\phi}{|\nabla\phi|}.
\end{equation}	
Here it should be noted that the time derivative in Eq. (\ref{GoverPre}) has been artificially divided into two terms for the ease of handling in some works \cite{Rojas2015JCP,Ramirez2004PRE,Cartalade2016CMA}.
To give a simple form and to account for the influence of melt convection, in this work the following modified governing equations for the solute and heat transfer in the whole region are adopted,
\begin{subequations}
	\begin{equation}\label{GoverC}
		\frac{\partial C}{\partial t}+\nabla\cdot \left(C\mathbf{u}\right)=\left(1-k\right)\nabla\cdot\left[D\tilde{q}\left(\phi\right)\nabla U-\mathbf{J}_{at}\right],
	\end{equation}
	\begin{equation}\label{GoverT}
		\frac{\partial H}{\partial t}+\nabla\cdot \left(H\mathbf{u}\right)=\alpha\nabla^2\theta,
	\end{equation}
\end{subequations}
where $C=c/c_{\infty}$ is the dimensionless concentration, $H=\theta-h\left(\phi\right)/2$, and $\mathbf{u}$ is the velocity. 

In addition, when the melt flow and the fluid-solid interaction are considered, the following NS equations for incompressible Newtonian flows must be considered,
\begin{subequations}
	\begin{equation}
		\nabla\cdot\mathbf{u}=0,
	\end{equation}
	\begin{equation}
		\frac{\partial\mathbf{u}}{\partial t}+\nabla\cdot\left(\mathbf{uu}\right)=-\nabla p+\nabla\cdot\nu\nabla\mathbf{u}+\mathbf{F}+\left[1-\tilde{q}\left(\phi\right)\right]\mathbf{f},
	\end{equation}
\end{subequations}
where $p$ is the pressure, $\nu$ is viscosity, $\mathbf{F}$ is the external force, $\mathbf{f}$ is the force caused by the fluid-solid interaction which is limited by the interpolation function $\left[1-\tilde{q}\left(\phi\right)\right]$. The expression of $\mathbf{f}$ is given in subsection \ref{LBMflow}, which is different from the dissipative drag force in Refs. \cite{Beckermann1999JCP,Rojas2015JCP} where an empirical constant $h^*=2.757$ is used.

\section{The diffuse-interface lattice Boltzmann method for dendritic growth}\label{LBMs}
In this section, we will propose a diffuse-interface MRT-LB method, which is composed of four different LB models for above governing equations used to describe the dendritic growth. For brevity, we first present a unified LB model for the convection-diffusion type equation and NS equations.

Similar to Ref. \cite{Chai2020PRE}, the semidiscrete LB evolution equation can be written as
\begin{equation}\label{LBE}
	f_i\left(\mathbf{x}+\mathbf{c}_i\Delta t,t+a_n\Delta t\right)=f_i\left(\mathbf{x},t\right)-\Lambda_{ij}\left(f_j-f_j^{eq}\right)\left(\mathbf{x},t\right)+\Delta t\left(F_i+\frac{\Delta t}{2}\hat{D}_iF_i\right)\left(\mathbf{x},t\right)+\Delta t\left(\delta_{ij}-\frac{\Lambda_{ij}}{2}\right)G_j\left(\mathbf{x},t\right),
\end{equation}
where $f_i\left(\mathbf{x},t\right)$ ($i=0,1,\cdots,q-1$, $q$ represents the number of directions of the discrete velocity $\mathbf{c}_i$) is the distribution function of the macroscopic variable $\Psi$ at position $\mathbf{x}$ and time $t$, $f_i^{eq}\left(\mathbf{x},t\right)$ is the corresponding equilibrium distribution function. $F_i\left(\mathbf{x},t\right)$ and $G_i\left(\mathbf{x},t\right)$ are the distribution functions of source/force term and auxiliary term, respectively. $\Lambda_{ij}$ is the collision matrix, $\hat{D}_i=a_n\partial_t+\gamma\mathbf{c}_i\cdot\nabla$ with $\gamma\in\{0,1\}$ being a parameter, $a_n$ is the relaxation of the time step $\Delta t$.

Through choosing some specific moments of distribution functions, one can get the LB models for the macroscopic governing equations through the direct Taylor expansion, and some details are shown in \ref{DTE}. In the following, some remarks are listed.

\begin{remark}
	If $a_n=1$, Eq.\,(\ref{LBE}) would degenerate into the standard LB model \cite{Chai2020PRE}.
\end{remark}
\begin{remark}\label{remarkPF}
	For the phase-field equation with $a_n=a_s^2\left(\mathbf{n}\right)$, the distribution function on the left-hand side of Eq.\,(\ref{LBE}) can be approximated by $f_i\left(\mathbf{x}+\mathbf{c}_i\Delta t,t+a_n\Delta t\right)=a_nf_i\left(\mathbf{x}+\mathbf{c}_i\Delta t,t+\Delta t\right)+\left(1-a_n\right)f_i\left(\mathbf{x}+\mathbf{c}_i\Delta t,t\right)$. 
	Additionally, for the nonlinear source term $Q_{\phi}$, the term of $\Delta t^2\hat{D}_iF_i/2$ can be simplified by $\Delta t^2a_s^2\left(\mathbf{n}\right)\partial_tF_i/2$ with $\gamma=0$.
\end{remark}
\begin{remark}
	For the general convection-diffusion equation with a simple source term, the term $\hat{D}_iF_i$ in Eq.\,(\ref{LBE}) can be discretized by $\hat{D}_iF_i\left(\mathbf{x},t\right)=\left[F_i\left(\mathbf{x}+\mathbf{c}_i\Delta t,t+a_n\Delta t\right)-F_i\left(\mathbf{x},t\right)\right]/\Delta t$ under the condition of $\gamma=1$. In this case, one can obtain the following evolution equation through introducing the new variable $\bar{f}_i=f_i-\Delta tF_i/2$,
	\begin{equation}
			\bar{f}_i\left(\mathbf{x}+\mathbf{c}_i\Delta t,t+a_n\Delta t\right)=\bar{f}_i\left(\mathbf{x},t\right)-\Lambda_{ij}\left(\bar{f}_j-f_j^{eq}\right)\left(\mathbf{x},t\right)+\Delta t\left(\delta_{ij}-\frac{\Lambda_{ij}}{2}\right)\left[F_j\left(\mathbf{x},t\right)+G_j\left(\mathbf{x},t\right)\right].
		\end{equation}
	The macroscopic variable should be calculated by
	\begin{equation}
			\Psi=\sum_if_i=\sum_i\bar{f}_i+\frac{\Delta t}{2}\sum_iF_i.
		\end{equation}
\end{remark}

\subsection{Aniostropic lattice Boltzmann model for phase field} 
According to Eq. (\ref{LBE}) and Remark \ref{remarkPF}, the evolution equation of LB model for phase field can be expressed as
\begin{align}\label{LBE-phi}
	\begin{autobreak}
		a_s^2\left(\mathbf{n}\right)f_i\left(\mathbf{x}+\mathbf{c}_i\Delta t,t+\Delta t\right)=
		f_i\left(\mathbf{x},t\right)-\left[1-a_s^2\left(\mathbf{n}\right)\right]f_i\left(\mathbf{x}+\mathbf{c}_i\Delta t,t\right)
		-\Lambda_{ij}^{\phi}\left[f_j\left(\mathbf{x},t\right)-f_j^{eq}\left(\mathbf{x},t\right)\right]
		+\Delta tF_{i}^{\phi}\left(\mathbf{x},t\right)
		+\frac{\Delta t^2}{2}a_s^2\left(\mathbf{n}\right)\partial_tF_i^{\phi}\left(\mathbf{x},t\right)
		+\Delta t\left(\delta_{ij}-\frac{\Lambda_{ij}^{\phi}}{2}\right)G_j^{\phi}\left(\mathbf{x},t\right).
	\end{autobreak}	
\end{align}
Here the equilibrium, source and auxiliary distribution functions appeared in above equation are given by
\begin{equation}
	f_i^{eq}=\omega_i\phi,\quad F_i^{\phi}=\omega_i\frac{Q_{\phi}}{\tau_0},\quad G_i^{\phi}=-\frac{\omega_i\mathbf{c}_i\cdot \mathbf{N}}{a_s^2\left(\mathbf{n}\right)},
\end{equation}
where $\omega_i$ is the weight coefficient. It should be noted that the term $W_0^2\nabla\cdot\mathbf{N}$ in Eq. (\ref{GoverPhi}) is regarded as a source term appeared in $G_i^{\phi}$, instead of the convection term in the previous works \cite{Cartalade2016CMA,Younsi2016JCP,Sun2019IJHMT,Wang2020CMS,Wang2021CMAME}. And also, to correctly recover Eq. (\ref{GoverPhi}), the term $\Delta t\partial_t\sum_i\mathbf{c}_if_i$ in Chapman–Enskog/Taylor expansion is neglected in these available works \cite{Cartalade2016CMA,Younsi2016JCP,Sun2019IJHMT,Wang2020CMS,Wang2021CMAME}, which may cause some unexpected errors. In addition, the term $\Delta t^2a_s^2\left(\mathbf{n}\right)\partial_tF_i^{\phi}$ is added in the evolution equation to obtain the phase-field equation (\ref{GoverPhi}) at the order of $O(\Delta t^2)$.

The macroscopic order parameter is calculated by zero-order moment of distribution function,
\begin{equation}\label{macro-phi}
	\phi=\sum_if_i.
\end{equation}
Finally, the gradient of order parameter $\nabla\phi$, as another important variable, is also needed in the simulation of phase field for dendritic growth, and how to calculate it has a significant influence on the accuracy and stability of the numerical method. Actually, following the previous works \cite{Chai2013PRE,Chai2014PRE,Wang2021CMAME}, we can obtain a local computing scheme of $\nabla\phi$ from Eq. (\ref{localDphi}) in the LB framework,
\begin{equation}
	\nabla\phi=-\frac{s_1^{\phi}}{\hat{c}_s^2\Delta t}\left(\sum_i\mathbf{c}_if_i+\frac{W_0^2\mathbf{N}}{\tau_0}\right),
\end{equation}
where $\hat{c}_s$ is the lattice sound speed, $s_1^{\phi}=1/\left[a_s^2\left(\mathbf{n}\right)W_0^2/\tau_0\hat{c}_s^2\Delta t+0.5\right]$ is the relaxation parameter. 

\subsection{Lattice Boltzmann model for solute transfer}
The convection-diffusion equation (\ref{GoverC}) for the solute transfer can be first rewritten as the following form with constant coefficient in the diffusive flux, 
\begin{equation}\label{GoverC1}
	\frac{\partial C}{\partial t}+\nabla\cdot\left(C\mathbf{u}\right)=D\left(1-k\right)\nabla\cdot\left\{\nabla \left[\tilde{q}\left(\phi\right)U\right]-U\nabla \tilde{q}\left(\phi\right)-\frac{\mathbf{J}_{at}}{D}\right\},
\end{equation}
and the evolution equation of LB model for Eq. (\ref{GoverC1}) can be given by
\begin{align}\label{LBE-C}
	\begin{autobreak}
		n_i\left(\mathbf{x}+\mathbf{c}_i\Delta t,t+\Delta t\right)=
		n_i\left(\mathbf{x},t\right)
		-\Lambda_{ij}^C\left[n_j\left(\mathbf{x},t\right)-n_j^{eq}\left(\mathbf{x},t\right)\right]+\Delta t\left(\delta_{ij}-\frac{\Lambda_{ij}^C}{2}\right)G_j^C\left(\mathbf{x},t\right).
	\end{autobreak}	
\end{align}
To correctly recover the solute transfer equation, the distribution functions are defined as
\begin{equation}
	n_i^{eq}=\begin{cases}
		C+\left(\omega_i-1\right)\tilde{q}\left(\phi\right)U, &i=0\\
		\omega_i\tilde{q}\left(\phi\right)U+\omega_i\mathbf{c}_i\cdot C\mathbf{u}/\hat{c}_s^2, &i\neq 0
	\end{cases},\quad G_i^C=\omega_i\mathbf{c}_i\cdot\left[\frac{\partial_t\left(C\mathbf{u}\right)}{\hat{c}_s^2}+U\nabla \tilde{q}\left(\phi\right)+\frac{\mathbf{J}_{at}}{D}\right].
\end{equation}

The dimensionless concentration and supersaturation are computed from the following relations,
\begin{equation}\label{macro-C}
	C=\sum_in_i,\quad U=\frac{\frac{2C}{1+k-\left(1-k\right)\phi}-1}{1-k}.
\end{equation}

\subsection{Lattice Boltzmann model for heat transfer}
Similarly,  when the heat transfer is considered, the evolution equation of LB model for the convection-diffusion equation (\ref{GoverT}) can be expressed as
\begin{align}\label{LBE-T}
	\begin{autobreak}
		g_i\left(\mathbf{x}+\mathbf{c}_i\Delta t,t+\Delta t\right)=
		g_i\left(\mathbf{x},t\right)
		-\Lambda_{ij}^{\theta}\left[g_j\left(\mathbf{x},t\right)-g_j^{eq}\left(\mathbf{x},t\right)\right]+\Delta t\left(\delta_{ij}-\frac{\Lambda_{ij}^{\theta}}{2}\right)G_j^{\theta}\left(\mathbf{x},t\right).
	\end{autobreak}	
\end{align}
To obtain the correct Eq. (\ref{GoverT}), the distribution functions appeared in above equation should be given by
\begin{equation}
	g_i^{eq}=\begin{cases}
		\omega_i\theta-h\left(\phi\right)/2, &i=0\\
		\omega_i\theta+\omega_i\mathbf{c}_i\cdot H\mathbf{u}/\hat{c}_s^2, &i\neq0
	\end{cases},\quad 
	G_i^{\theta}=\frac{\omega_i\mathbf{c}_i\cdot\partial_t\left(H\mathbf{u}\right)}{\hat{c}_s^2}.
\end{equation}

The macroscopic temperature is calculated by
\begin{equation}\label{macro-T}
	\theta=\sum_ig_i+\frac{h\left(\phi\right)}{2}.
\end{equation} 

\subsection{Diffuse interface lattice Boltzmann model for melt flow}\label{LBMflow}
The evolution equation of LB model for flow field reads
\begin{align}\label{LBE-u}
	\begin{autobreak}
		h_i\left(\mathbf{x}+\mathbf{c}_i\Delta t,t+\Delta t\right)=
		h_i\left(\mathbf{x},t\right)
		-\Lambda_{ij}^{\mathbf{u}}\left[h_j\left(\mathbf{x},t\right)-h_j^{eq}\left(\mathbf{x},t\right)\right]+\Delta t\left(\delta_{ij}-\frac{\Lambda_{ij}^{\mathbf{u}}}{2}\right)F_j^{\mathbf{u}}\left(\mathbf{x},t\right).
	\end{autobreak}
\end{align}

To derive the incompressible NS equations with fluid-solid interaction, the distribution functions in Eq. (\ref{LBE-u}) are designed as
\begin{equation}
	h_i^{eq}=\sigma_i+\omega_i\left[\frac{\mathbf{c}_i\cdot\mathbf{u}}{\hat{c}_s^2}+\frac{\mathbf{uu}:\left(\mathbf{c}_i\mathbf{c}_i-\hat{c}_s^2\mathbf{I}\right)}{2\hat{c}_s^4}\right],\quad F_i^{\mathbf{u}}=\omega_i\left\{\mathbf{c}_i\cdot\frac{\mathbf{F}+\left[1-\tilde{q}\left(\phi\right)\right]\mathbf{f}}{\hat{c}_s^2}+\varphi\frac{\left(\mathbf{uF}+\mathbf{Fu}\right):\left(\mathbf{c}_i\mathbf{c}_i-\hat{c}_s^2\mathbf{I}\right)}{2\hat{c}_s^4}\right\},
\end{equation}
where $\sigma_0=\left(\omega_0-1\right)p/\hat{c}_s^2+\rho_0$, $\sigma_i=\omega_ip/\hat{c}_s^2 \left(i\neq 0\right)$ \cite{He2004CP}, $\rho_0$ is a constant representing the density of the pure fluid, and $\varphi$ is an adjustable parameter.
 
The macroscopic velocity and pressure are determined by
\begin{equation}\label{macro-u}
	\mathbf{u}^*=\sum_i\mathbf{c}_ih_i+\frac{\Delta t}{2}\mathbf{F},\quad \mathbf{u}=\mathbf{u}^*+\frac{\Delta t}{2}\left[1-\tilde{q}\left(\phi\right)\right]\mathbf{f},\quad p=\frac{\hat{c}_s^2}{1-\omega_0}\left(\sum_{i\neq0}h_i-\omega_0\frac{\mathbf{u}\cdot\mathbf{u}}{2\hat{c}_s^2}+\varphi\Delta tK\frac{\mathbf{u}\cdot\mathbf{F}}{\hat{c}_s^2}\right),
\end{equation}
where $\mathbf{u}^*$ is the velocity without considering the fluid-solid interaction, $\mathbf{u}$ is the corrected velocity. $K$ is a parameter related to the relaxation parameters (see details in \ref{D2Q59}). Here the second order term of $\mathbf{f}$ in $F_i^{\mathbf{u}}$ \cite{Liu2022CF} is neglected for simplicity. We would also like to point out that the fluid-solid interaction $\mathbf{f}$ can be discretized as $\left(\mathbf{u}_s-\mathbf{u}^*\right)/\Delta t$ with $\mathbf{u}_s$ being the velocity of solid point, which is similar to the forcing scheme in the immersed boundary method \cite{Kang2011IJNMF} and smoothed profile method \cite{Jafari2011PRE,Mino2017PRE}.

In addition, the hydrodynamic force $\mathbf{F}_f$ and torque $T_f$ can be calculated by
\begin{subequations}\label{HDforce}
	\begin{equation}
		\mathbf{F}_f=-\left(\Delta x\right)^d\sum_n\mathbf{f}\left(\mathbf{x}_n\right),
	\end{equation}
	\begin{equation}
		T_f=-\left(\Delta x\right)^d\sum_n\left(\mathbf{x}_n-\mathbf{x}_s\right)\times\mathbf{f}\left(\mathbf{x}_n\right),
	\end{equation}
\end{subequations}
where $\Delta x$ is the lattice spacing, $\mathbf{x}_n$ is the coordinate of the nodes covered by the solid, and $\mathbf{x}_s$ is the centroid of the solid phase.

\section{Numerical validations and discussion}\label{Numerical}
In this section, several two-dimensional (2D) problems are used to test the present diffuse-interface LB method. Specifically, the flow around a stationary circular cylinder is first adopted to show the capacity of the diffuse interface method in describing the fluid-solid interaction, then we considered the thermal dendritic growth, solutal dendritic growth as well as the thermosolutal dendritic growth with pure diffusion and melt flow, respectively. The MRT-LB models with D2Q5 and D2Q9 lattice structures are applied for convection-diffusion type equation and the NS equations, and the related transformation matrices as well as some moments of the distribution functions are shown in \ref{D2Q59}. The relaxation parameters related to diffusivities and viscosity are given by $s_1^C=1/\left[D\left(1-k\right)/\hat{c}_s^2\Delta t+0.5\right]$, $s_1^{\theta}=1/\left(\alpha/\hat{c}_s^2\Delta t+0.5\right)$ and $s_2^{\mathbf{u}}=1/\left(\nu/\hat{c}_s^2\Delta t+0.5\right)$, while the other relaxation parameters in the collision matrices are set to be 1 if not specified. The adjustable parameter $\varphi$ is fixed as 0 for simplicity. In our simulations, the cylinder or solid seed with radius $R_s$ is initialized as a diffuse circle: $\phi\left(\mathbf{x},0\right)=\tanh\left[\left(R_s-d_s\right)/\sqrt{2W_0}\right]$, where $d_s$ is the distance from the center of circle. The non-equilibrium extrapolation scheme \cite{Guo2002CP} is used to treat the physical boundary conditions imposed on the surround walls. Additionally, due to the large difference in the values of diffusivities and viscosity, two-time-scaling strategy is adopted \cite{Wang2021CMAME}, i.e., $\Delta t'=\Delta t/N$ for the simulation of melt flow, $N$ is a scale coefficient and set to be 15 in this work.

\subsection{The flow around a stationary circular cylinder}\label{cylinder}
The problem of flow around a stationary circular cylinder is first considered to test the diffuse-interface LB method in the computation of the hydrodynamic force. The flow is driven by a constant velocity $\mathbf{u}_{in}=(u_0,0)$ at inlet, and a free outflow boundary condition is imposed on the outlet boundary. To depict the flow pattern of this problem, the Reynolds number $Re=2u_0R_s/\nu$ is used. 

\begin{figure}
	\centering
	\subfigure[]{
		\begin{minipage}{0.49\linewidth}
			\centering
			\includegraphics[width=3.0in]{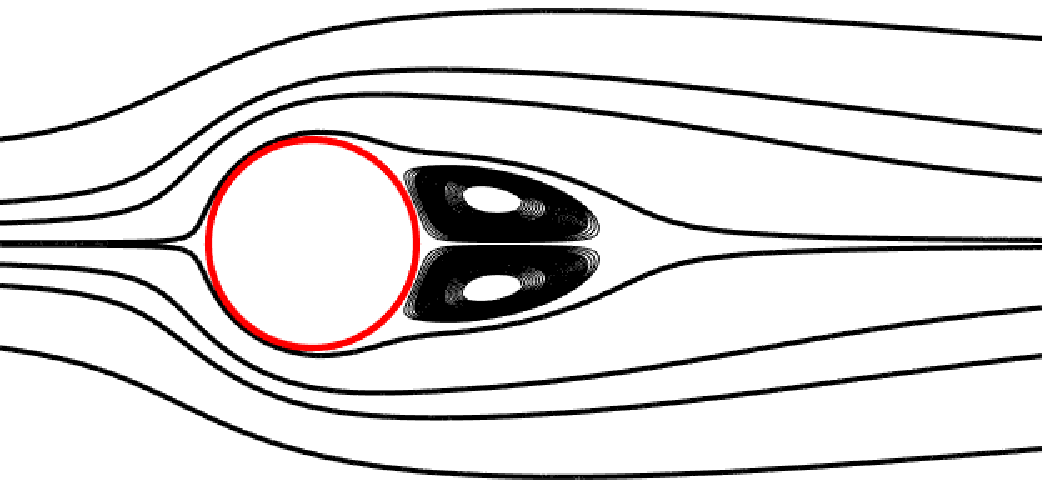}
			\label{fig-cylinder20}
	\end{minipage}}	
	\subfigure[]{
		\begin{minipage}{0.49\linewidth}
			\centering
			\includegraphics[width=3.0in]{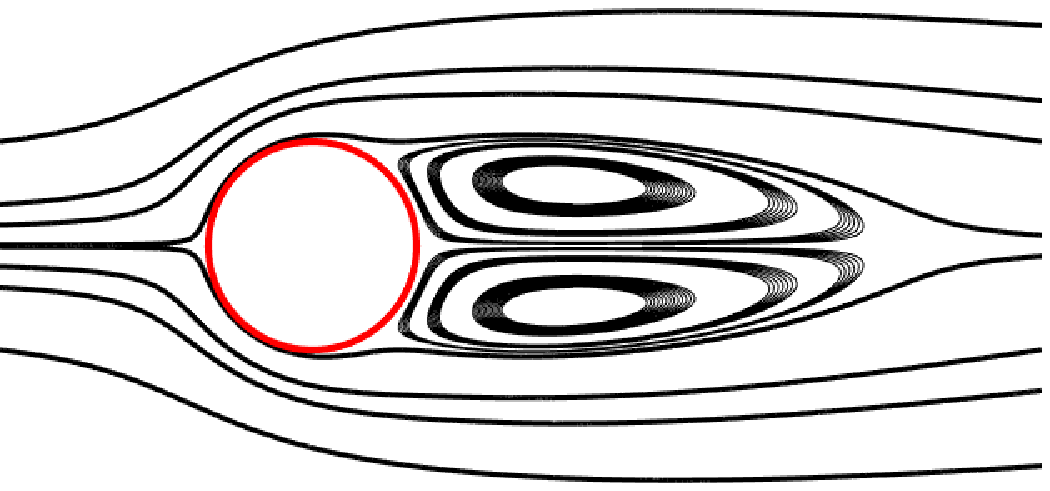}
			\label{fig-cylinder40}
	\end{minipage}}	
	\caption{The streamlines of the fluid flows around the stationary circular cylinder at $Re=20$ (a) and $Re=40$ (b).}
	\label{fig-cylinder}
\end{figure}

In our simulations, the computational domain is $70R_s\times40R_s$ with a uniform mesh size $1400\times800$, the circular cylinder with $R_s=0.5$ is placed at $(20R_s,20R_s)$, and the other parameters are set as $u_0=0.1$, $W_0=0.5\Delta x$ and $s_2^{\mathbf{u}}=1.25$. We performed some simulations with $Re=20$ and $Re=40$, and plotted the streamlines in Fig. \ref{fig-cylinder}. From this figure, one can observe that when the flow reaches steady state, a pair of symmetric recirculating eddies formed behind the cylinder, and the length of the recirculating region increases with the increase of $Re$. These results are qualitatively consistent with the previous works \cite{Niu2006PLA,Liu2022CF}. To give a quantitative comparison, the drag coefficient $C_d=F_{fx}/\rho_0R_su_0^2$ ($F_{fx}$ is the hydrodynamic force in the $x$ direction) and the dimensionless recirculation length $L_W=L/R_s$ ($L$ is the recirculation length) are measured and listed in Table \ref{table-cylinder}. As seen from this table, the results of the present method are in good agreement with some available data \cite{He1997JCP,Niu2006PLA,Liu2022CF}.
\begin{table}
	\centering
	\caption{A comparison of the drag coefficient and recirculation length between the present work and some previous studies.}
	\begin{tabular}{ccccccc}
		\toprule
		$Re$ && Reference  && $C_d$  && $L_W$ \\
		\midrule
		20 && Present             && 2.153 && 1.853 \\
		&& He and Doolen, 1997 \cite{He1997JCP}  && 2.152 && 1.842 \\
		&& Niu et al., 2006    \cite{Niu2006PLA} && 2.144 && 1.890 \\
		&& Liu et al., 2022    \cite{Liu2022CF}  && 2.167 && 1.873 \\
		40 && Present             && 1.598 && 4.491 \\
		&& He and Doolen, 1997 \cite{He1997JCP}  && 1.499 && 4.490 \\
		&& Niu et al., 2006    \cite{Niu2006PLA} && 1.589 && 4.520 \\
		&& Liu et al., 2022    \cite{Liu2022CF}  && 1.613 && 4.540 \\
		\bottomrule
	\end{tabular}
	\label{table-cylinder}
\end{table}
\begin{figure}
	\centering
	\includegraphics[width=3.5in]{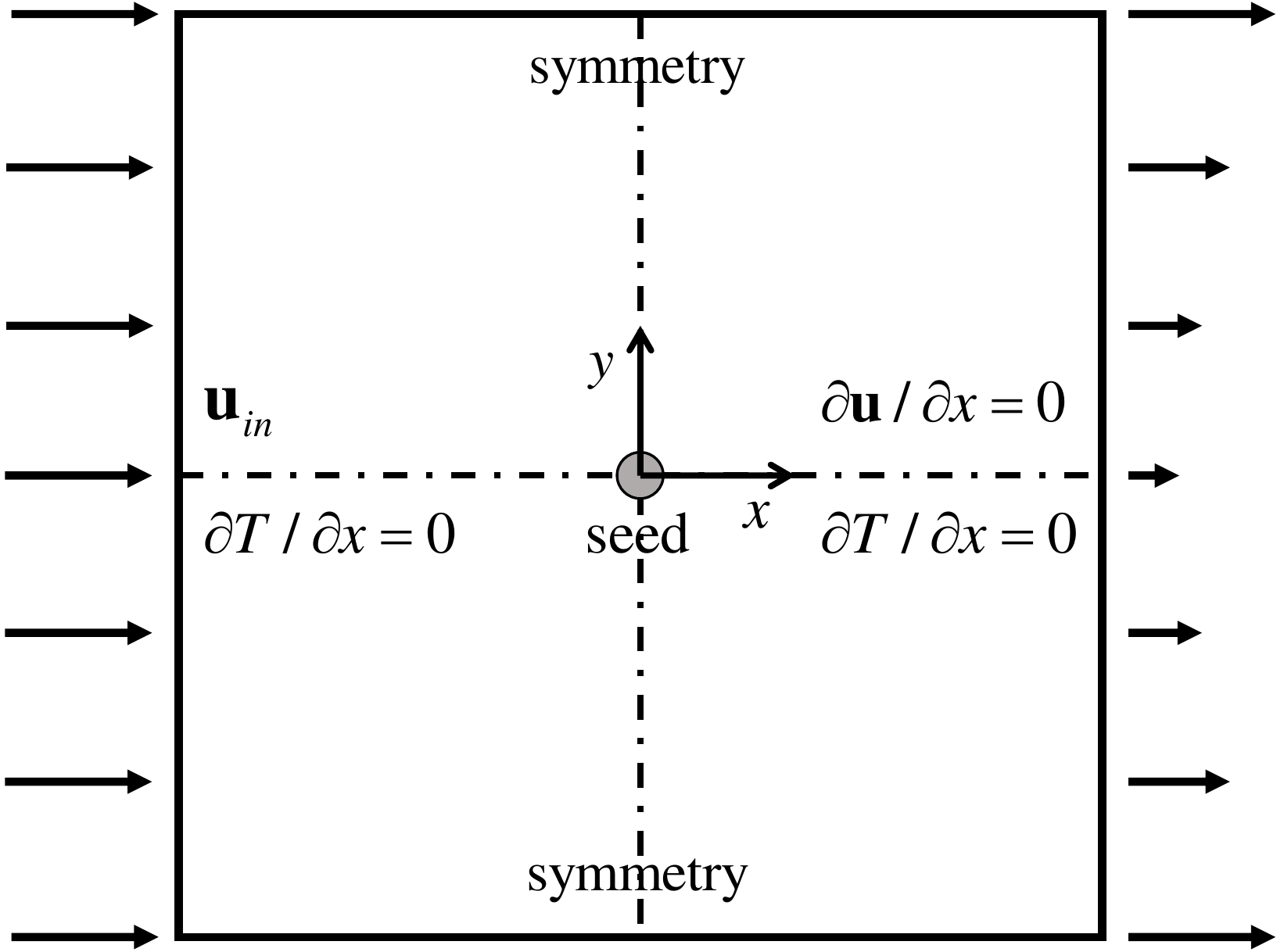}
	\caption{Initial and boundary conditions of the equiaxed dendritic growth with melt flow.}
	\label{fig-domain}
\end{figure}

\subsection{Thermal dendritic growth}\label{thermal}
We now consider the dendritic growth of a pure substance into a uniformly supercooled melt. Initially, a circular seed with $R_s=10\Delta x$ is placed at the center of the 2D square domain $L\times L$ \cite{Sun2019IJHMT}, and the initial temperature is set as a uniform value $\theta_0=-\Delta$. As shown in Fig. \ref{fig-domain}, for temperature field, the symmetry boundary condition is adopted on all walls, while for the flow field, $\mathbf{u}_{in}=\left(W_0/\tau_0,0\right)$ and $\partial\mathbf{u}/\partial x=\mathbf{0}$ are imposed on the left and right boundaries. It should be noted that for the pure diffusion case, $\mathbf{u}=\mathbf{0}$.  

In the following simulations, $L=512\Delta x$ with $\Delta x=0.4W_0$, $W_0=1$, $\Delta t=0.008\tau_0$, $\tau_0=1$, $\varepsilon_s=0.05$, $Mc_{\infty}=0$, $\Delta=0.55$, the initial supersaturation $U_0=0$ and $\alpha=4W_0^2/\tau_0$. The coupling coefficient is defined by $\lambda=a_1W_0/d_0$ with capillary length $d_0=0.1385W_0$ and $a_1=0.8839$. When the flow field is considered, the Prandtl number is given by $Pr=\nu/\alpha=23.1$. 

\begin{figure}
	\centering
	\subfigure[]{
		\begin{minipage}{0.49\linewidth}
			\centering
			\includegraphics[width=3.0in]{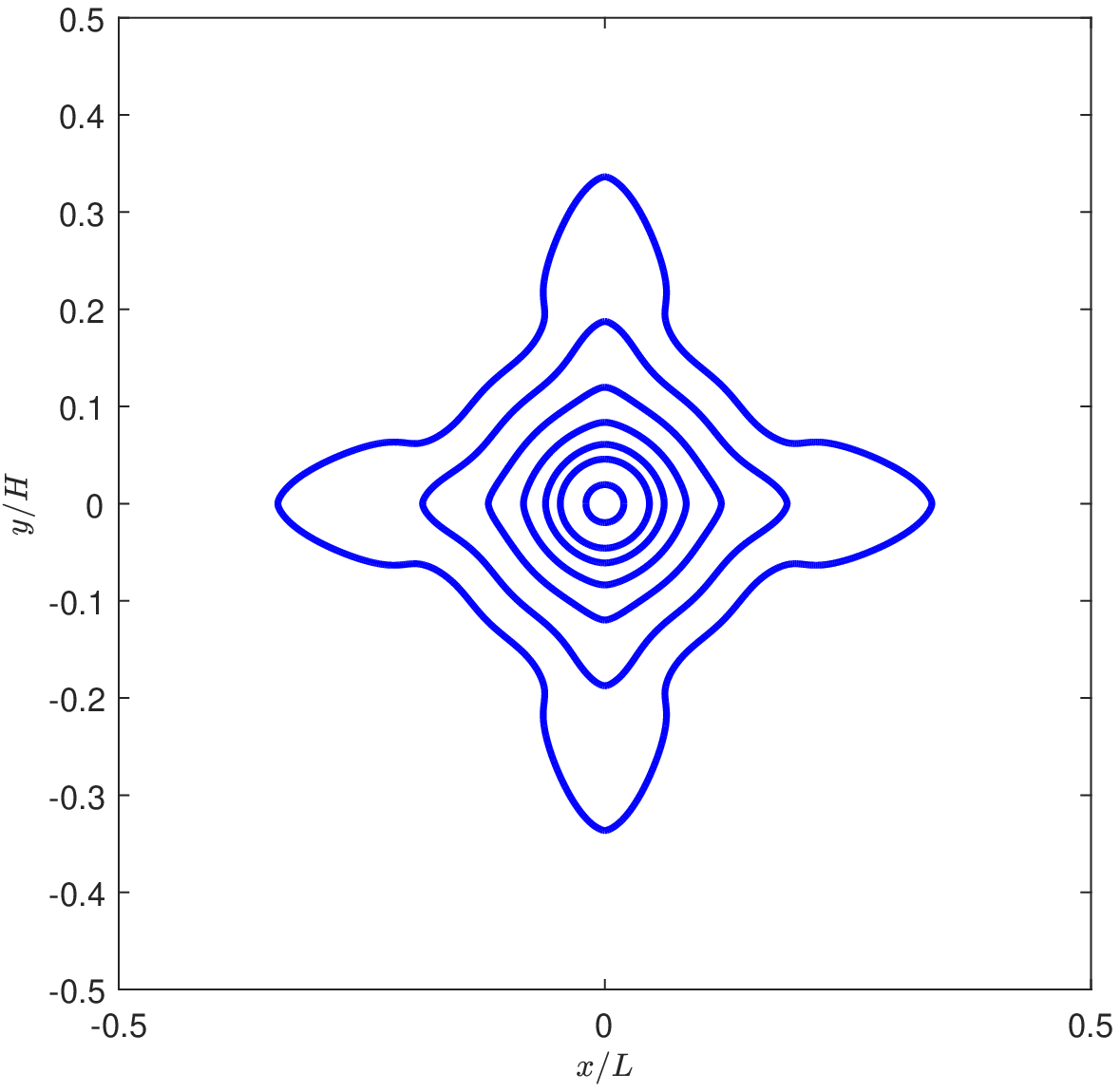}
			\label{fig-Tpure-phi}
	\end{minipage}}	
	\subfigure[]{
		\begin{minipage}{0.49\linewidth}
			\centering
			\includegraphics[width=3.0in]{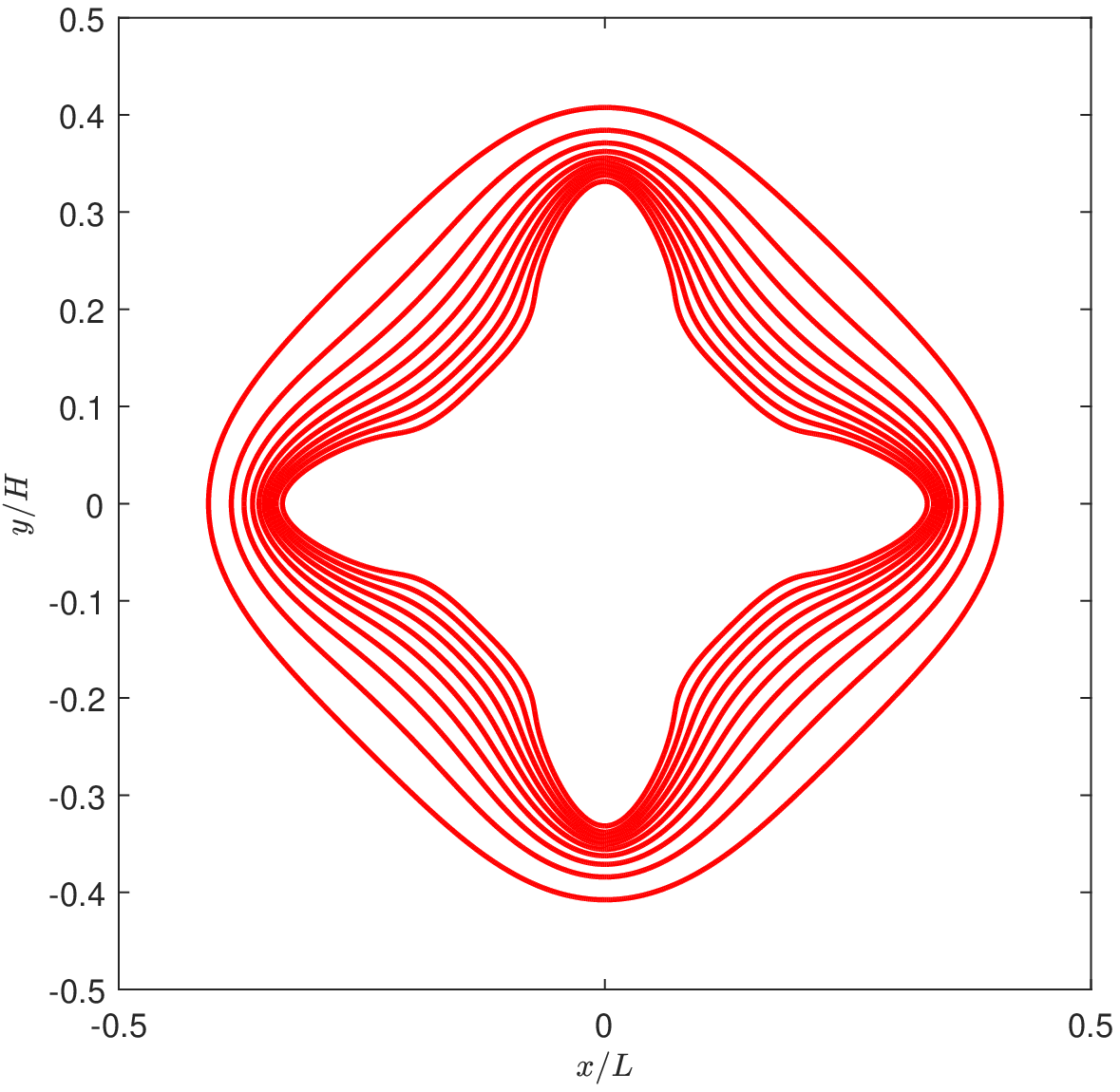}
			\label{fig-Tpure-T}
	\end{minipage}}	
	\caption{The interface evolution at $t/\tau_0=0,4,8,16,32,64,128$ (a) and isothermal lines from $\theta=-0.55$ to $\theta=-0.05$ with the increment of $0.05$ at $t/\tau_0=128$ (b) in the thermal denritic growth with pure diffusion.}
	\label{fig-Tpure}
\end{figure}

\begin{figure}
	\centering
	\subfigure[]{
		\begin{minipage}{0.49\linewidth}
			\centering
			\includegraphics[width=3.0in]{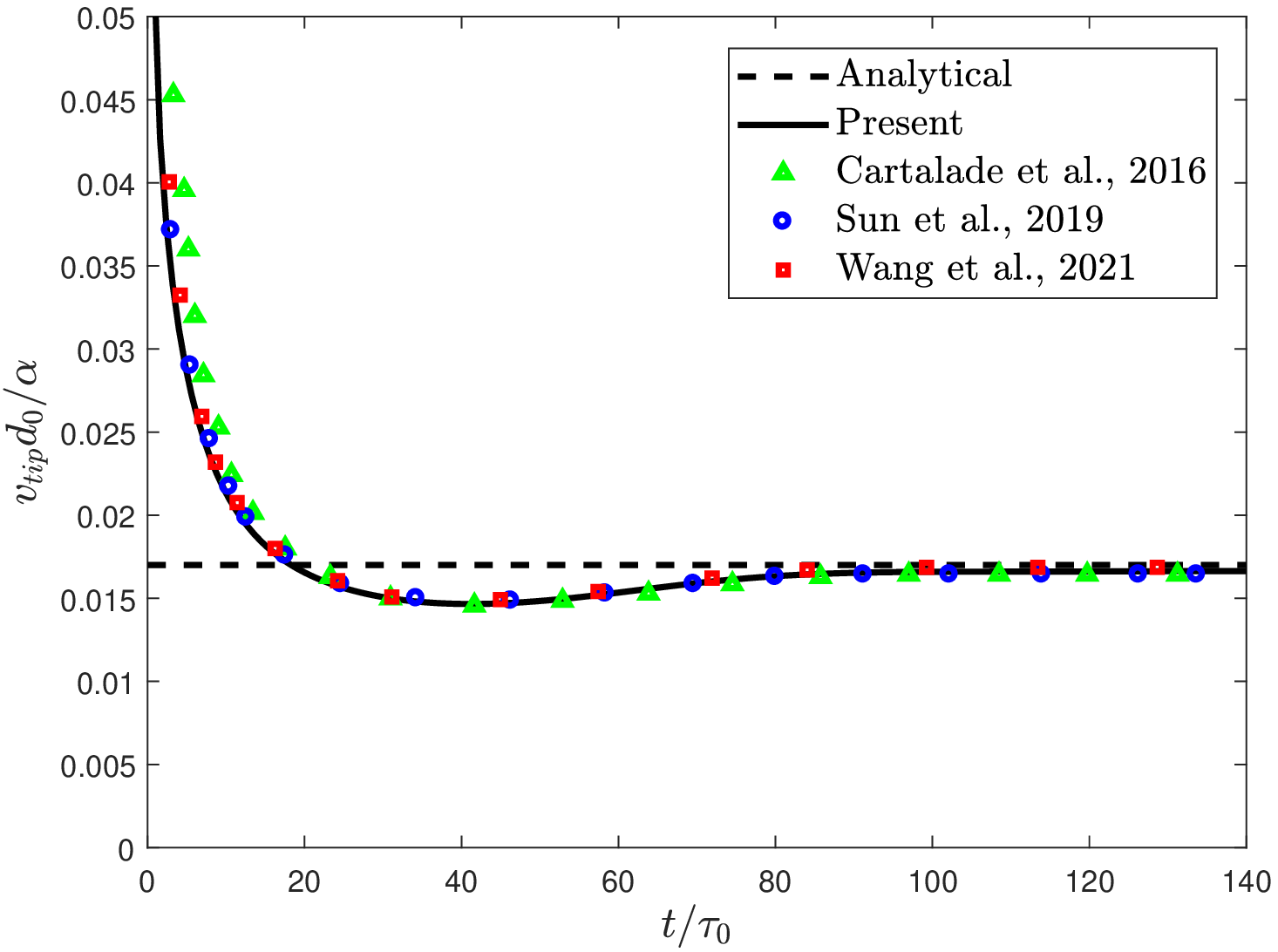}
			\label{fig-Tpure-Vtip}
	\end{minipage}}	
	\subfigure[]{
		\begin{minipage}{0.49\linewidth}
			\centering
			\includegraphics[width=3.0in]{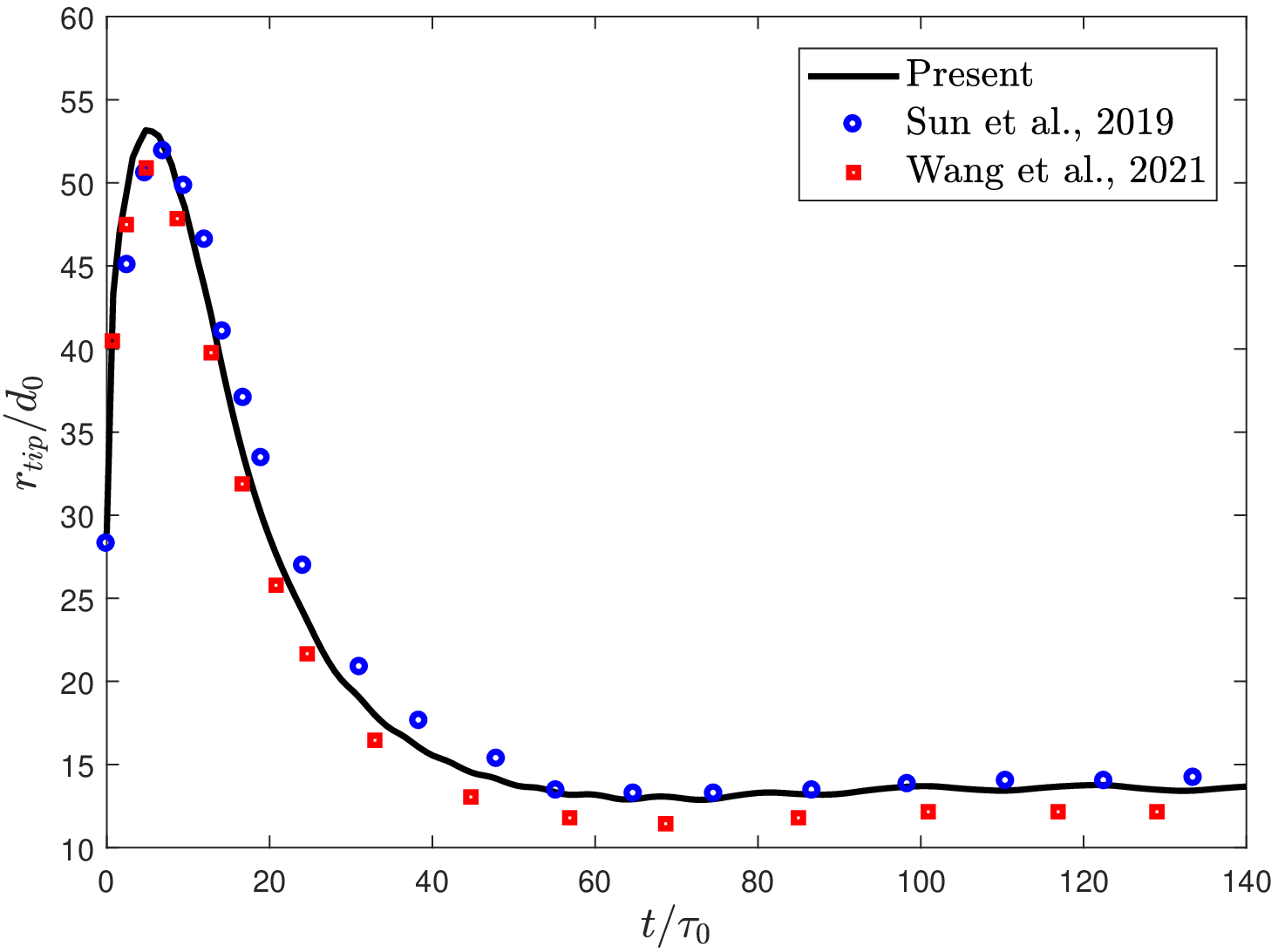}
			\label{fig-Tpure-Rtip}
	\end{minipage}}	
	\caption{Evolutions of tip velocity (a) and tip radius (b) in the thermal dendritic growth with pure diffusion.}
	\label{fig-Tpure-tip}
\end{figure}

In the study of dendritic growth with the pure diffusion, we presented the evolution of the interface and the isothermal lines at $t/\tau_0=128$ in Fig. \ref{fig-Tpure}. From this figure, one can find that the dendrite grows symmetrically with the fourfold anisotropy, and the gradients of temperature around the tips are much higher than those close to the dendritic root. These results agree well with some previous works \cite{Cartalade2016CMA,Sun2019IJHMT,Wang2021CMAME}. To further quantitatively describe the growth of dendrite, the tip velocity and tip radius are also measured in Fig. \ref{fig-Tpure-tip}. As shown in this figure, the tip velocity decreases rapidly, and with the increase of time, it reaches to a constant value $v_{tip}=0.0166$ that is very close to the Green's function analytical solution (0.017) \cite{Karma1996PRE} as well as some numerical results reported in Refs. \cite{Cartalade2016CMA,Sun2019IJHMT,Wang2021CMAME}. Additionally, the evolution of tip radius in time also agrees well with those in Refs. \cite{Sun2019IJHMT,Wang2021CMAME}.

\begin{figure}
	\centering
	\subfigure[]{
		\begin{minipage}{0.49\linewidth}
			\centering
			\includegraphics[width=3.0in]{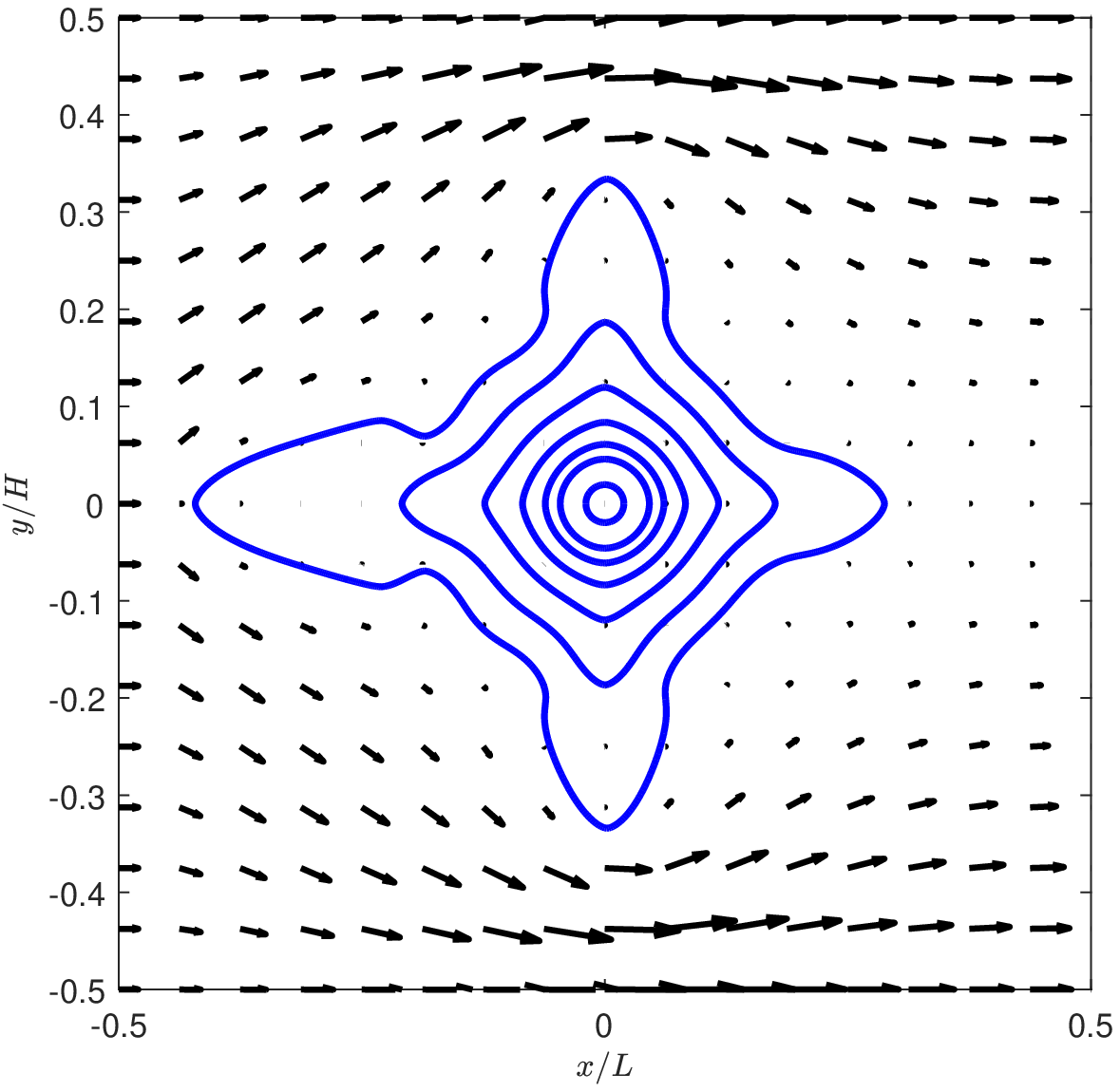}
			\label{fig-Tu-phi}
	\end{minipage}}	
	\subfigure[]{
		\begin{minipage}{0.49\linewidth}
			\centering
			\includegraphics[width=3.0in]{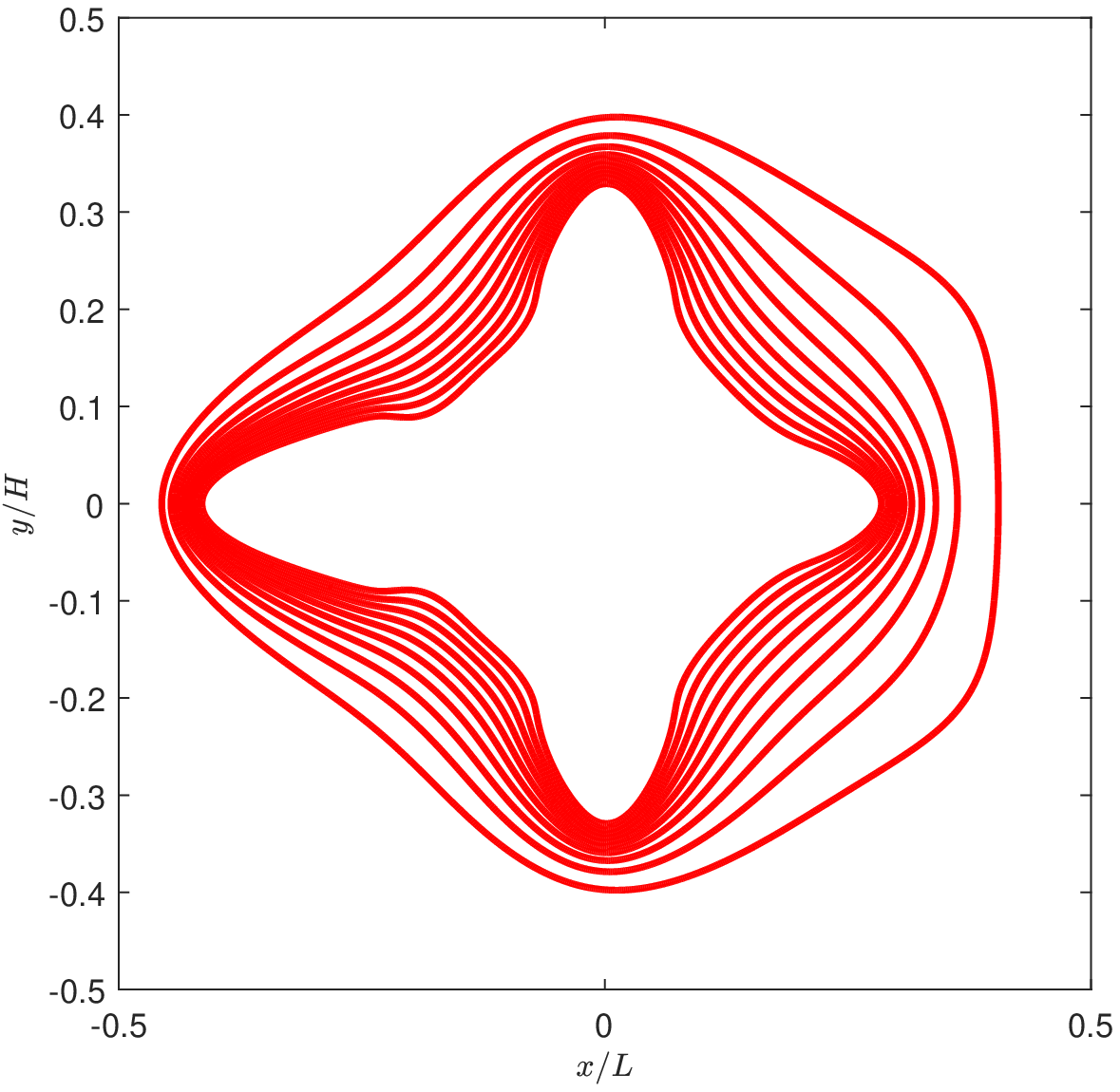}
			\label{fig-Tu-T}
	\end{minipage}}	
	\caption{The interface evolution at $t/\tau_0=0,4,8,16,32,64,128$ superimposed with velocity field at $t/\tau_0=128$ (a) and isothermal lines from $\theta=-0.5$ to $\theta=-0.05$ with the increment of $0.05$ at $t/\tau_0=128$ (b) in the thermal denritic growth with melt flow.}
	\label{fig-Tu}
\end{figure}

When the melt flow driven by the inlet velocity is considered, the dendritic growth displays a significant difference compared to the pure diffusion case. As seen from Fig. \ref{fig-Tu-phi}, the dendritic arm in the upstream direction is longer and thicker than the vertical and downstream arms. Besides, the temperature field is also affected by the melt flow, and the gradients of temperature around the upstream tip is larger than those around the vertical and downstream tips [see Fig. \ref{fig-Tu-T}]. 
With the increase of time, the velocity of upstream tip increases, the velocity of the vertical tips reaches a constant value which is the same as that of the pure diffusion case, while the velocity of downstream tip keeps to decrease and reaches a constant value. These results are consistent with those reported in Refs. \cite{Sun2019IJHMT,Wang2021CMAME}. In addition, it is also found that the evolutions of the tip radii are close to each other, and are also in agreement with some previous works.

\begin{figure}
	\centering
	\subfigure[]{
		\begin{minipage}{0.49\linewidth}
			\centering
			\includegraphics[width=3.0in]{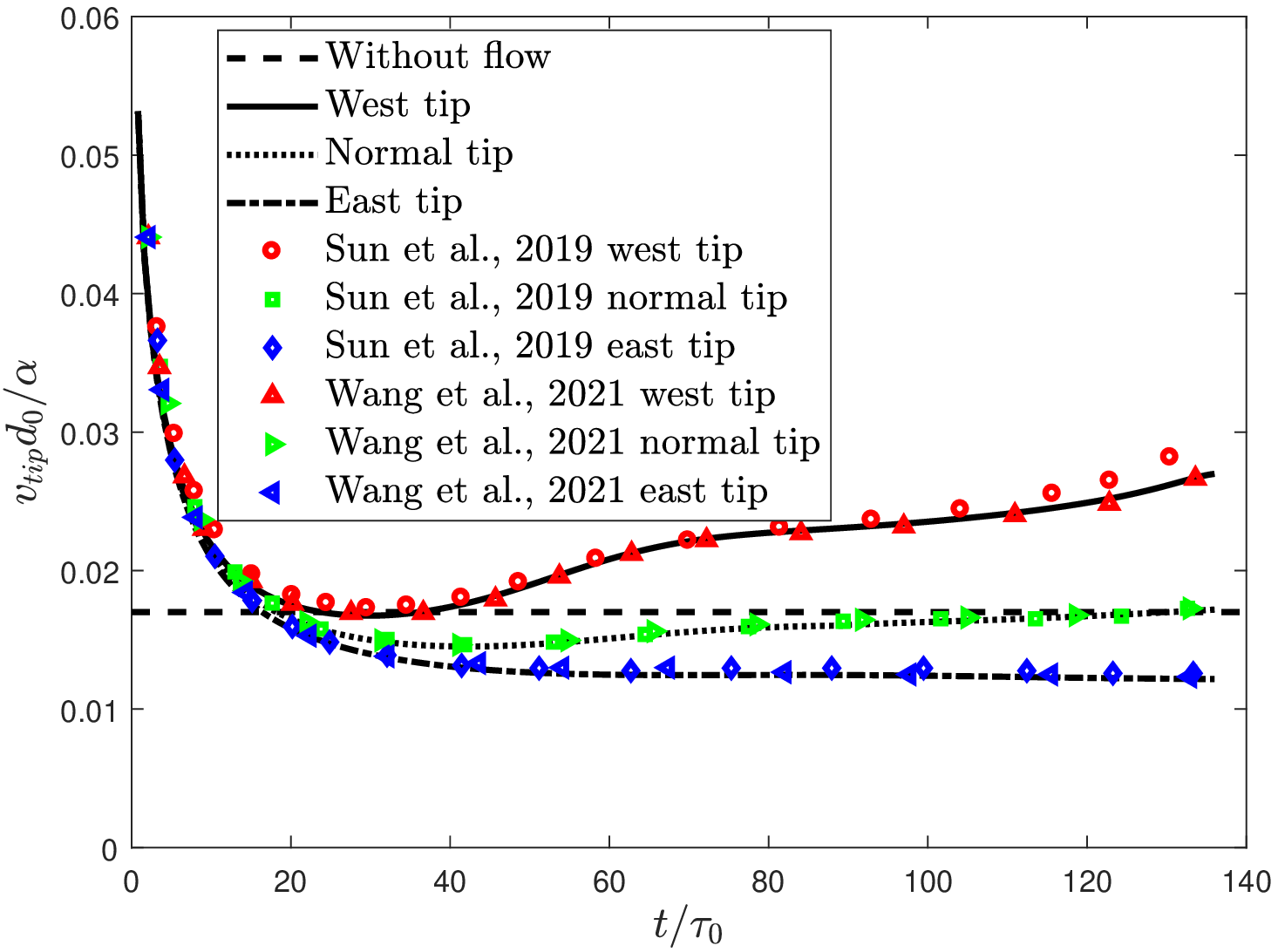}
			\label{fig-Tu-Vtip}
	\end{minipage}}	
	\subfigure[]{
		\begin{minipage}{0.49\linewidth}
			\centering
			\includegraphics[width=3.0in]{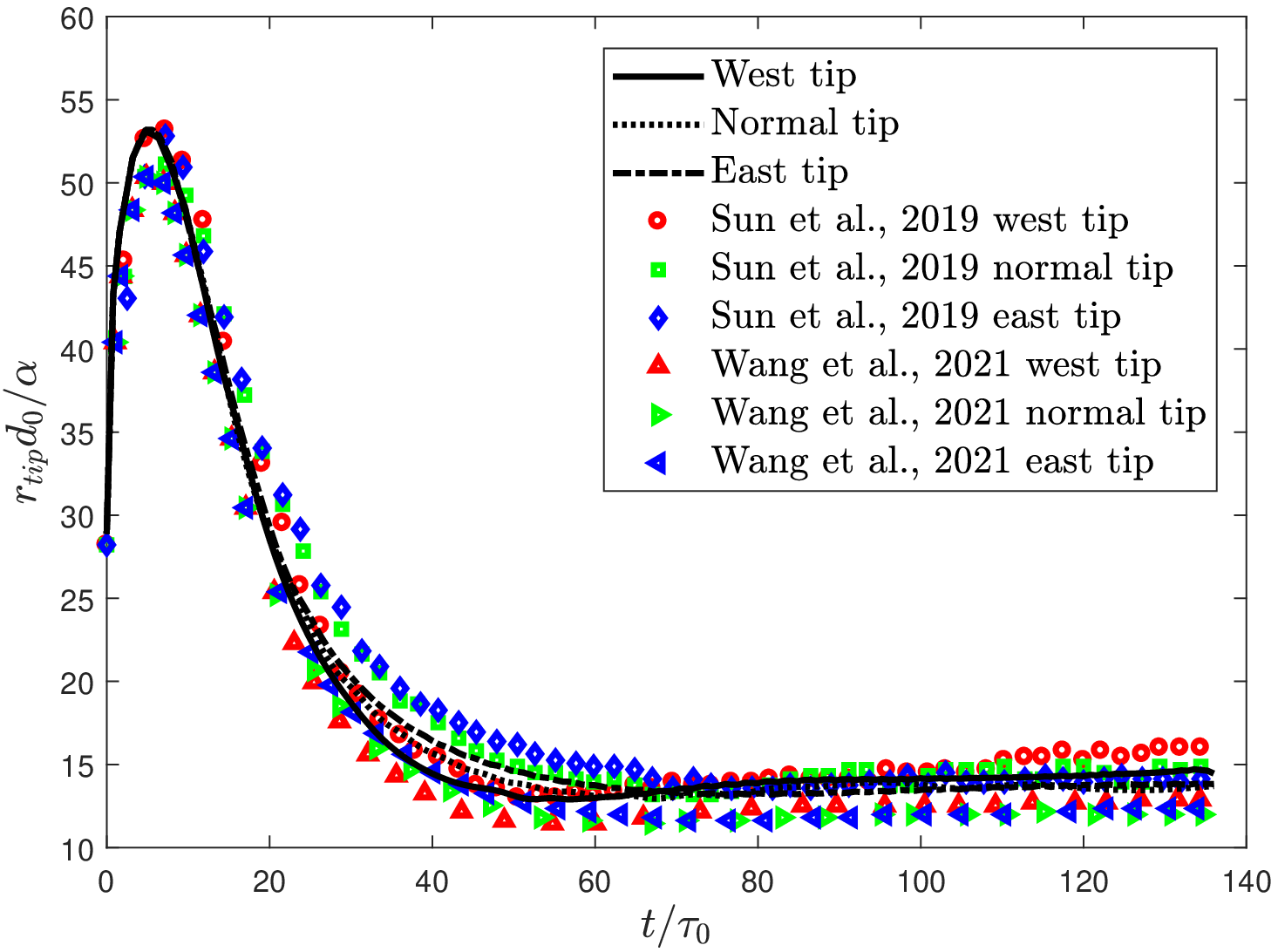}
			\label{fig-Tu-Rtip}
	\end{minipage}}	
	\caption{Evolutions of tip velocities (a) and tip radii (b) in the thermal dendritic growth with melt flow.}
	\label{fig-Tu-tip}
\end{figure}

\subsection{Solutal dendritic growth}\label{solutal}
\begin{figure}
	\centering
	\subfigure[]{
		\begin{minipage}{0.49\linewidth}
			\centering
			\includegraphics[width=3.0in]{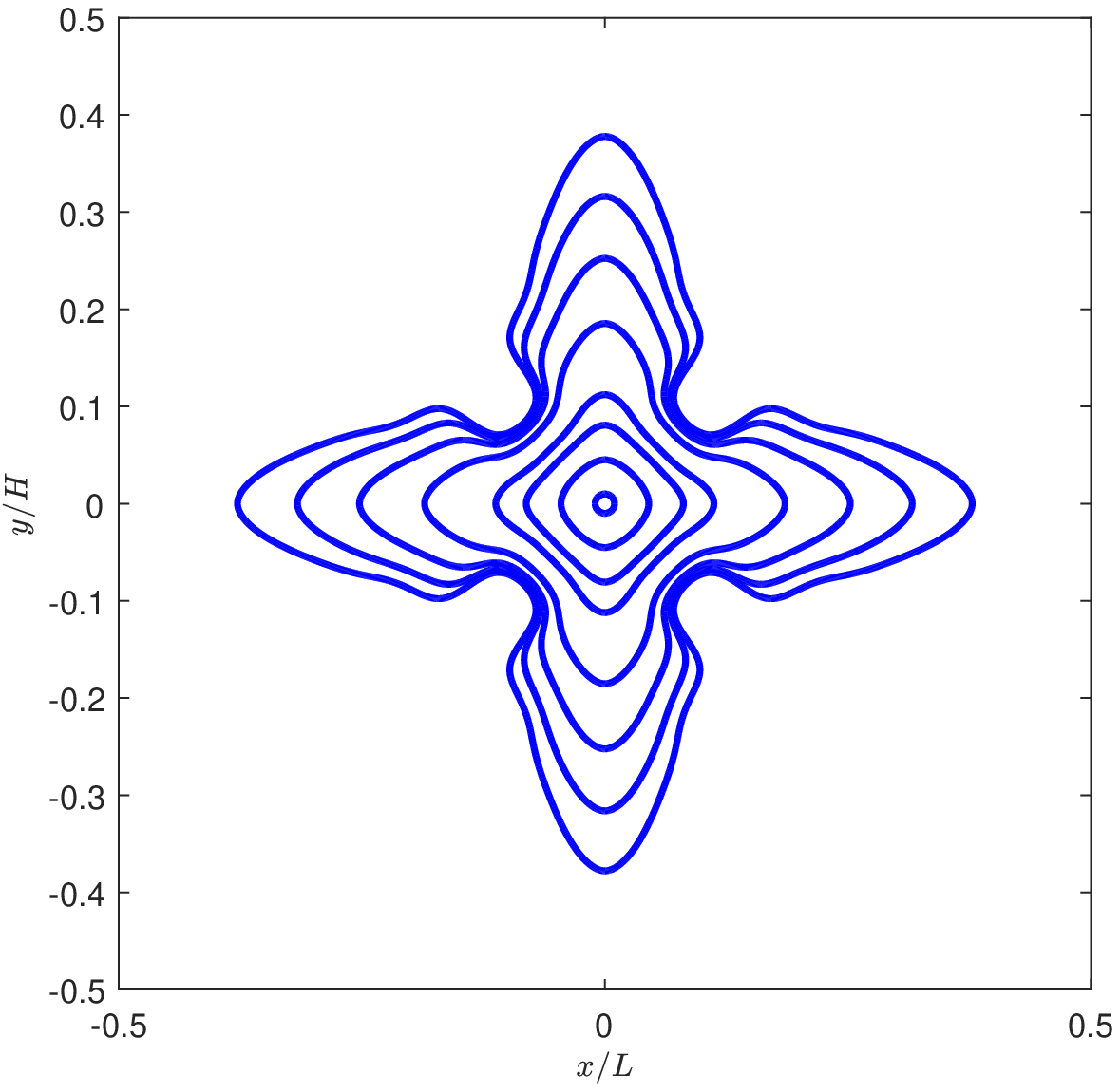}
			\label{fig-Cpure-phi}
	\end{minipage}}	
	\subfigure[]{
		\begin{minipage}{0.49\linewidth}
			\centering
			\includegraphics[width=3.0in]{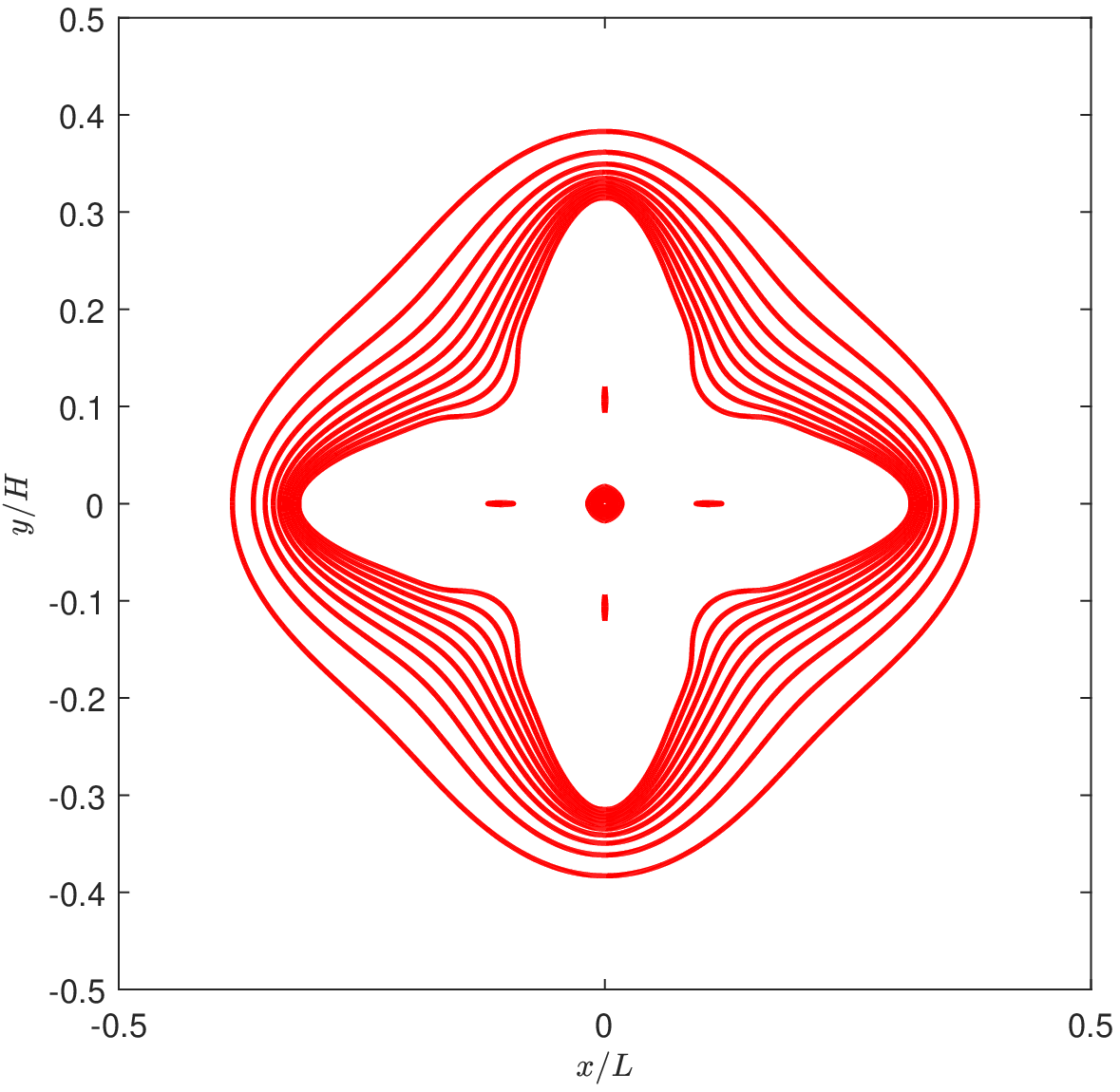}
			\label{fig-Cpure-T}
	\end{minipage}}	
	\caption{The interface evolution at $t/\tau_0=0,40,120,200,400,600,800,1000$ (a) and isosolutal lines from $U=-0.55$ to $U=-0.05$ with the increment of $0.05$ at $t/\tau_0=800$ (b) in the solutal denritic growth with pure diffusion.}
	\label{fig-Cpure}
\end{figure}
\begin{figure}
	\centering
	\subfigure[]{
		\begin{minipage}{0.49\linewidth}
			\centering
			\includegraphics[width=3.0in]{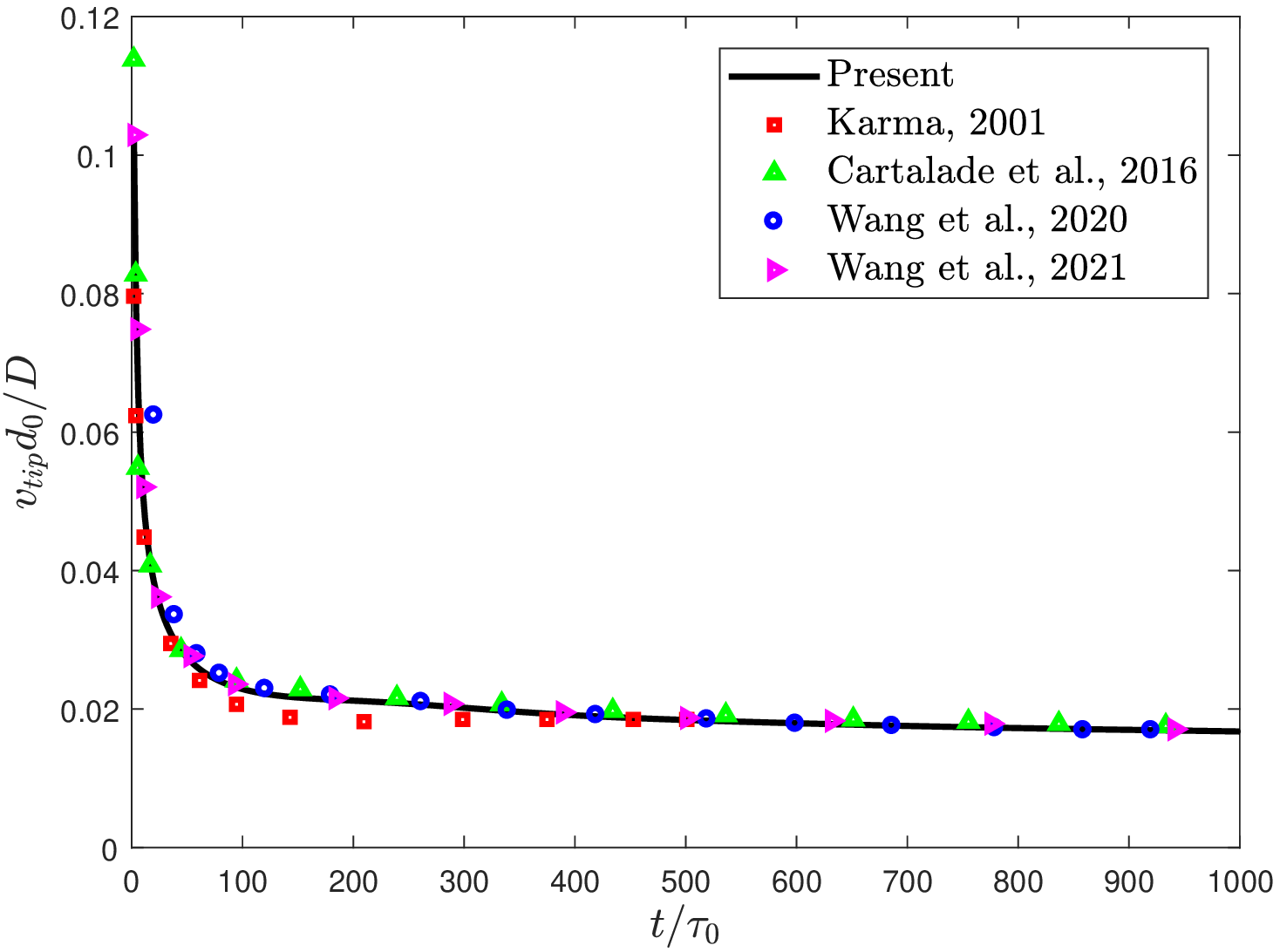}
			\label{fig-Cpure-Vtip}
	\end{minipage}}	
	\subfigure[]{
		\begin{minipage}{0.49\linewidth}
			\centering
			\includegraphics[width=3.0in]{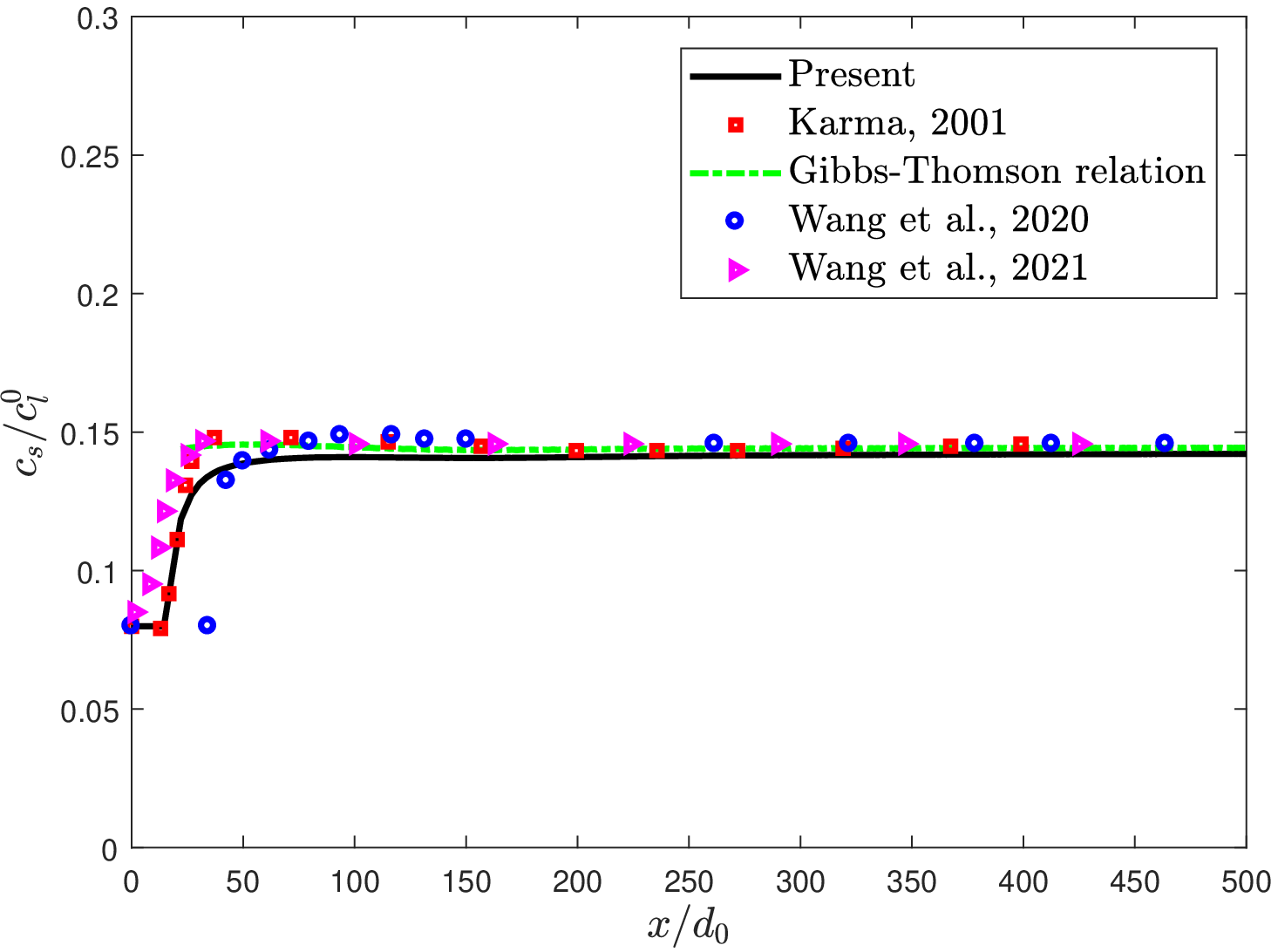}
			\label{fig-Cpure-Cs}
	\end{minipage}}	
	\caption{Evolution of tip velocity in the solutal dendritic growth with pure diffusion (a) and solute profiles on the solid side of interface along the central dendrite axis (b).}
	\label{fig-Cpure-tip}
\end{figure}
\begin{figure}
	\centering
	\subfigure[]{
		\begin{minipage}{0.49\linewidth}
			\centering
			\includegraphics[width=3.0in]{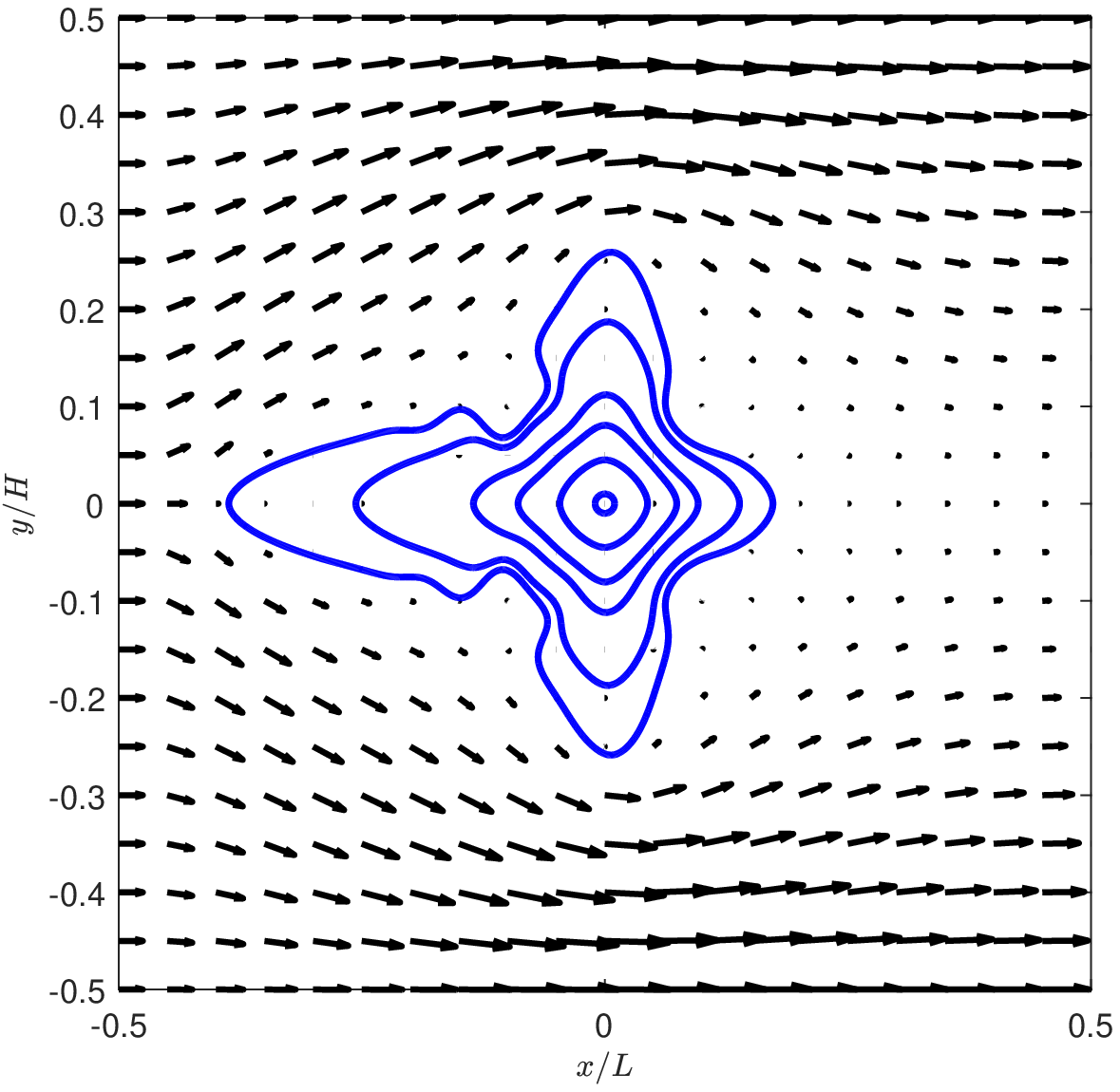}
			\label{fig-Cu-phi}
	\end{minipage}}	
	\subfigure[]{
		\begin{minipage}{0.49\linewidth}
			\centering
			\includegraphics[width=3.0in]{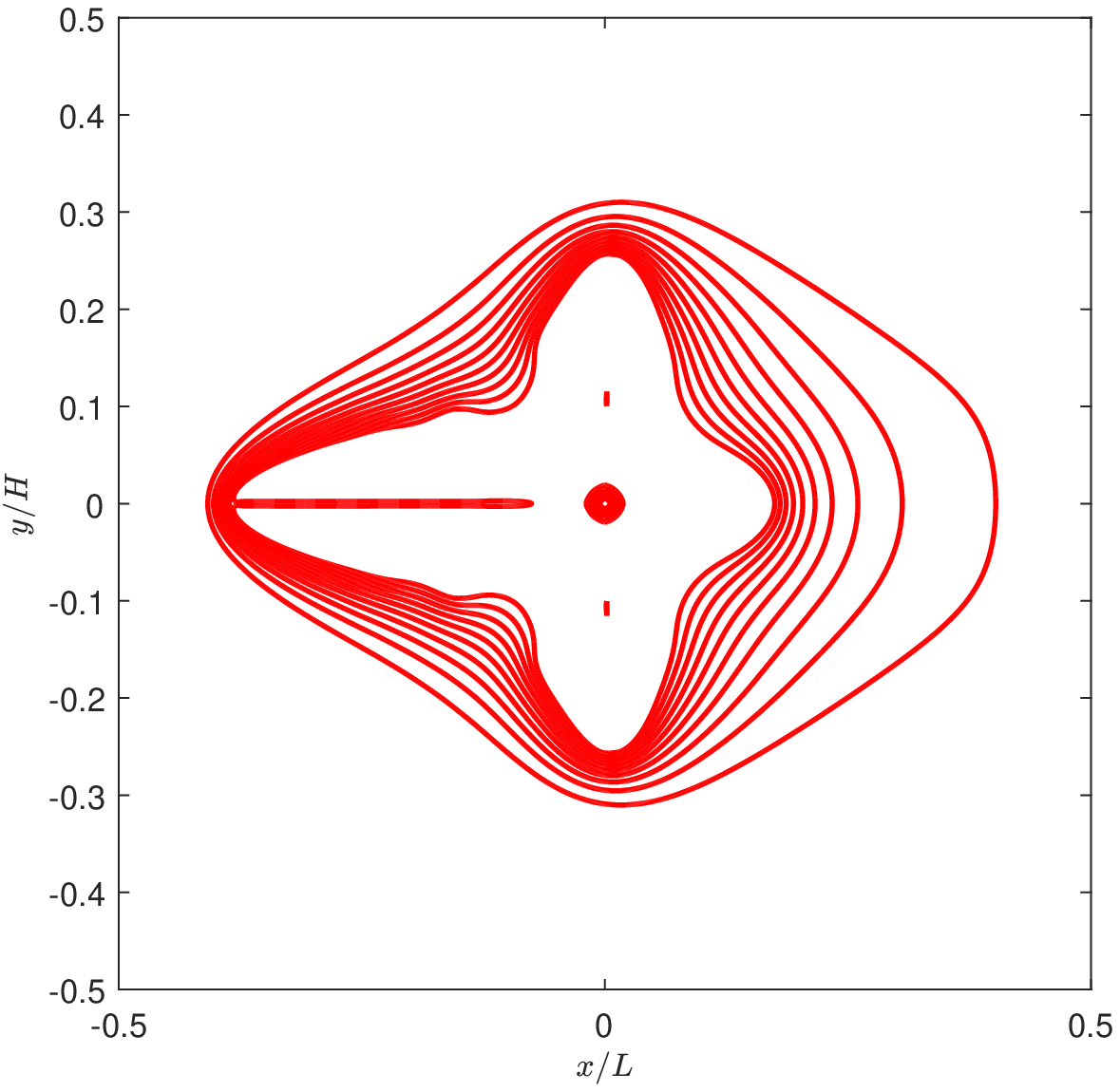}
			\label{fig-Cu-T}
	\end{minipage}}	
	\caption{The interface evolution at $t/\tau_0=0,40,120,200,400,600$ superimposed with velocity field at $t/\tau_0=600$ (a) and isosolutal lines from $U=-0.5$ to $U=-0.05$ with the increment of $0.05$ at $t/\tau_0=600$ (b) in the solutal denritic growth with melt flow.}
	\label{fig-Cu}
\end{figure}
\begin{figure}
	\centering
	\subfigure[]{
		\begin{minipage}{0.49\linewidth}
			\centering
			\includegraphics[width=3.0in]{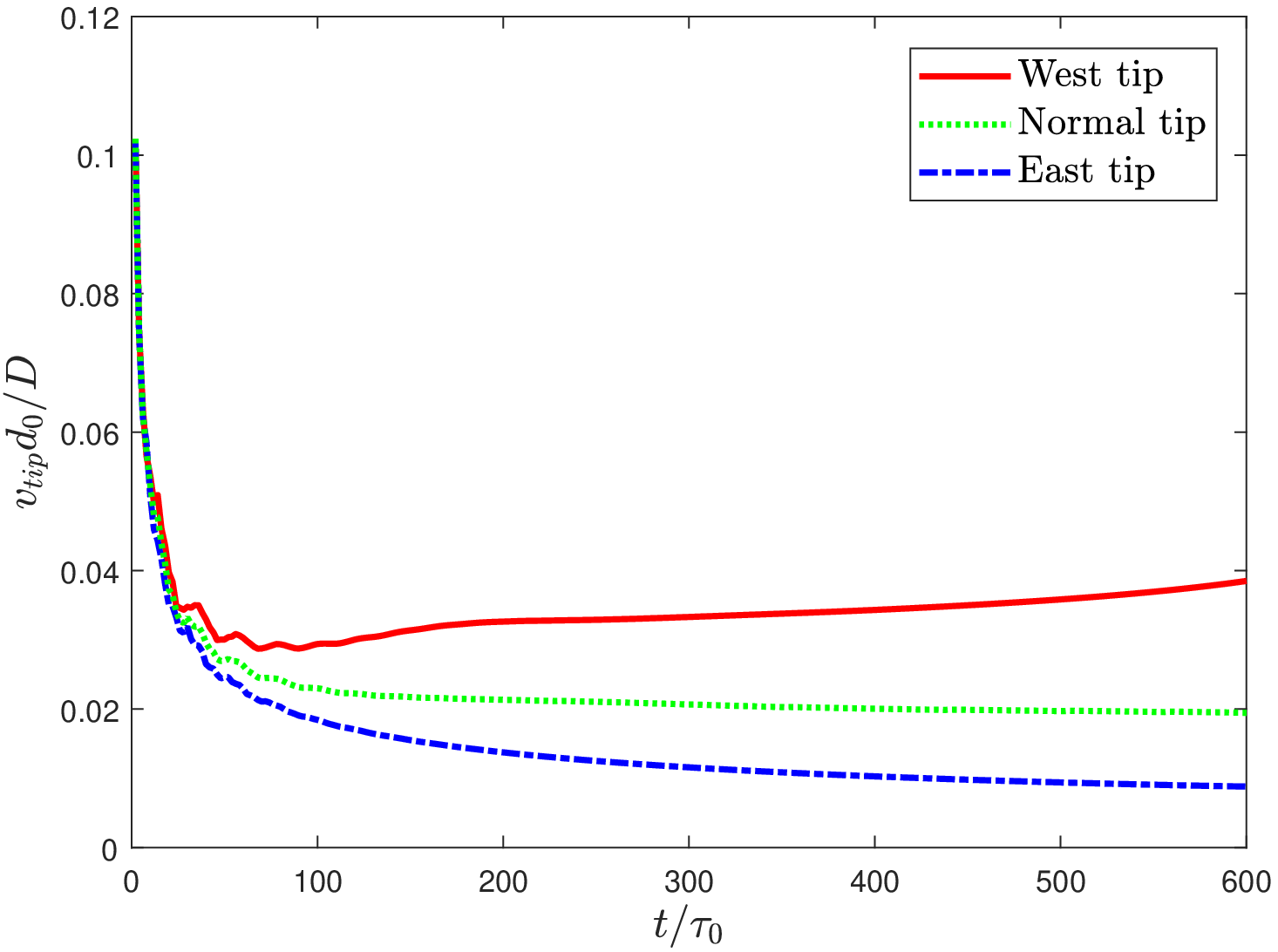}
			\label{fig-Cu-Vtip}
	\end{minipage}}	
	\subfigure[]{
		\begin{minipage}{0.49\linewidth}
			\centering
			\includegraphics[width=3.0in]{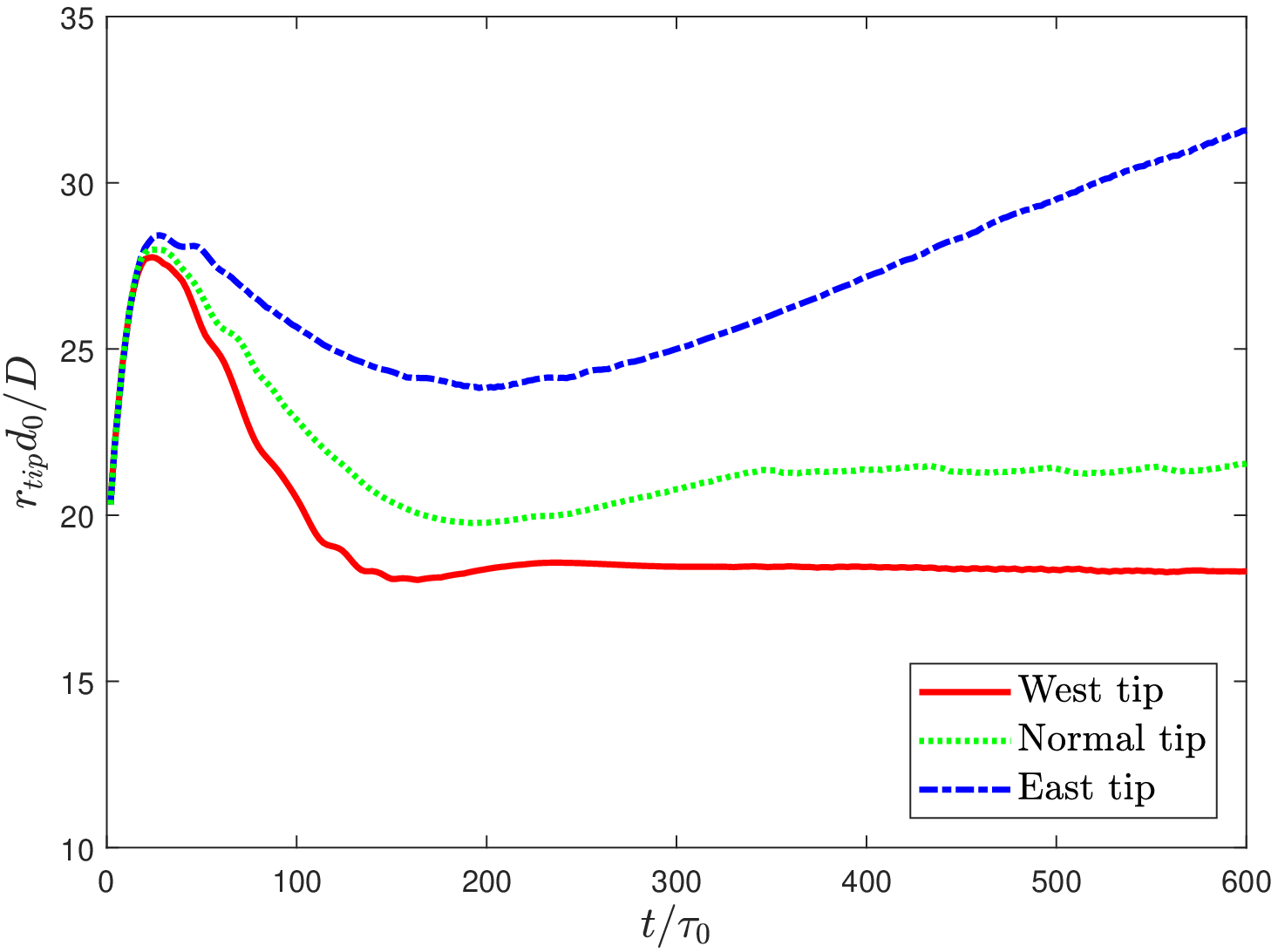}
			\label{fig-Cu-Rtip}
	\end{minipage}}	
	\caption{Evolutions of tip velocities (a) and tip radii (b) in the solutal dendritic growth with melt flow.}
	\label{fig-Cu-tip}
\end{figure}

For the dendritic growth of binary alloys under the isothermal condition, the computational domain and the circular seed are similar to those in subsection \ref{thermal}, and the symmetry boundary condition of supersaturation is adopted. In our simulations, the parameters are set as $L=1000\Delta x$, $\Delta t=0.02\tau_0$, $\varepsilon_s=0.02$, $\theta_0=0$, $U_0=-0.55$, $D=2$, $Mc_{\infty}=1$, $k=0.15$, $\lambda=3.191$ and $d_0=0.2762$. 

Figure \ref{fig-Cpure} presents the evolution of the interface morphology and the isosolutal lines at $t/\tau_0=800$, and these results are close to those in Refs. \cite{Wang2020CMS,Wang2021CMAME}. We also conducted some quantitative comparisons of tip velocity and the concentration on the solid side of interface versus displacement along the central dendritic axis. It is found that similar to the thermal case, the dendrite first grows at a relative high speed, then decreases dramatically, and finally reaches a steady state [see Fig. \ref{fig-Cpure-Vtip}]. The comparison of dimensionless concentration on the solid side of interface $c_s/c_l^0$ ($c_l^0$ is the equilibrium liquidus concentration) in Fig. \ref{fig-Cpure-Cs} also shows a good agreement with some available data \cite{Karma2001PRL,Wang2020CMS,Wang2021CMAME} and the Gibbs-Thomson relation $c_s/c_l^0=k\left[1-\left(1-k\right)d_0/r_{tip}\right]$, where $r_{tip}$ changes with the displacement along the central dendrite axis. It should be noted that the discrepancies near the dendrite center in Fig. \ref{fig-Cpure-Cs} are caused by the different initial seed radii and whether the diffusion of solute in solid is considered or not in the simulations.  

In addition, we also carried out some simulations of solutal dendritic growth with melt flow. From Fig. \ref{fig-Cu}, we can see that the upstream arm is much larger and thicker than the downstream one, which indicates that the melt flow plays a significant role on the dendritic growth. We also plotted the evolutions of tip velocities in Fig. \ref{fig-Cu-Vtip}, and it is found that the trends are similar to the thermal case with melt flow, but the velocities are relatively larger. On the other hand, the evolutions of tip radii have some obvious differences. All tip radii increase at the initial stage of the dendrite growth, then the upstream one decreases rapidly to a low value, but the vertical and downstream tips continue to increase after a small decrease. These phenomena may be caused by the more uniform supersaturation distribution around the vertical and downstream tips. 
  
\subsection{Thermosolutal dendritic growth}
\begin{figure}
	\centering
	\includegraphics[width=3.5in]{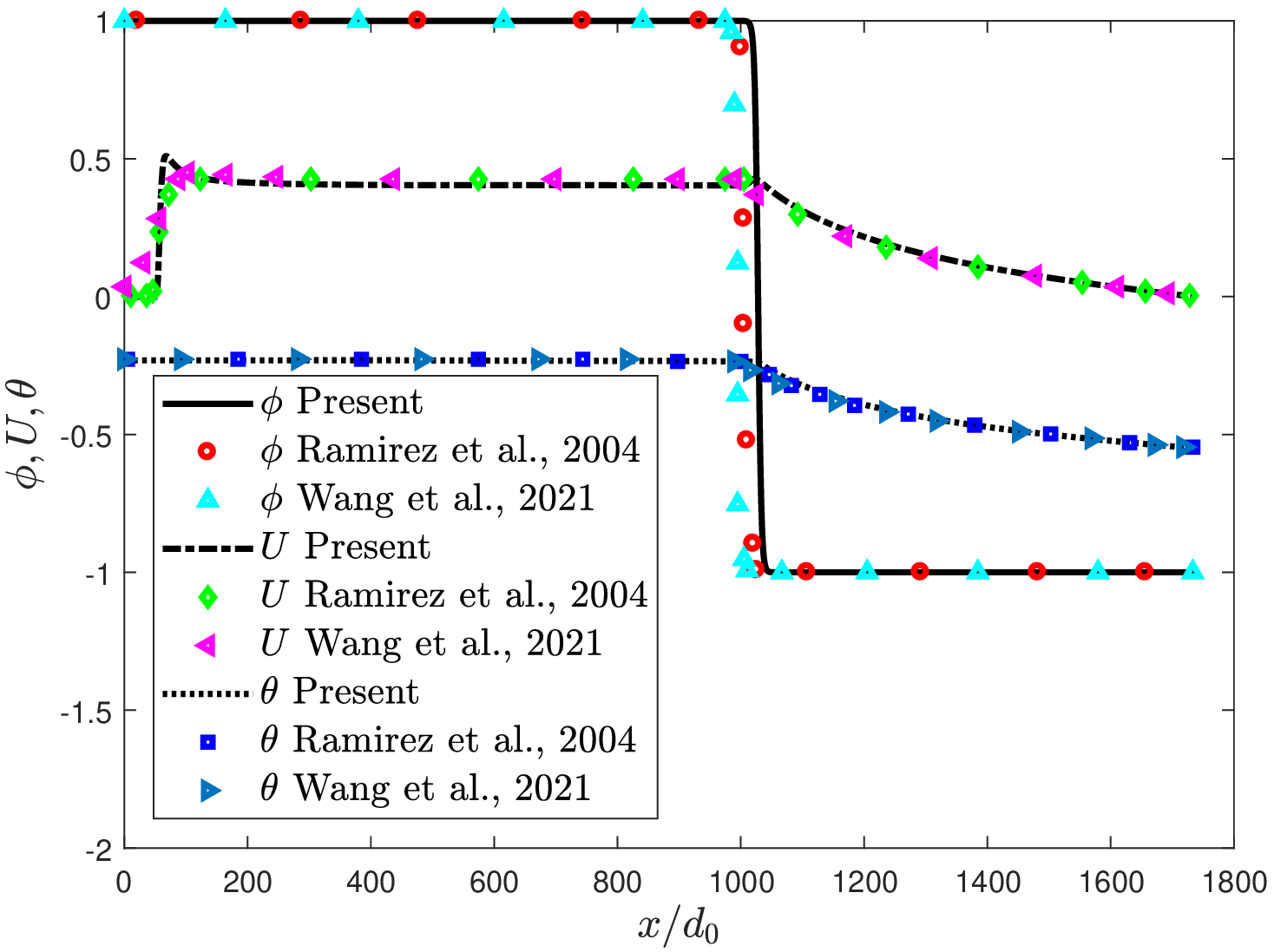}
	\caption{The profiles of $\phi$, $U$ and $\theta$ along the central dendrite axis at $tD/d_0^2=470000$.}
	\label{fig-TCpure}
\end{figure}
\begin{figure}
	\centering
	\subfigure[]{
		\begin{minipage}{0.49\linewidth}
			\centering
			\includegraphics[width=3.0in]{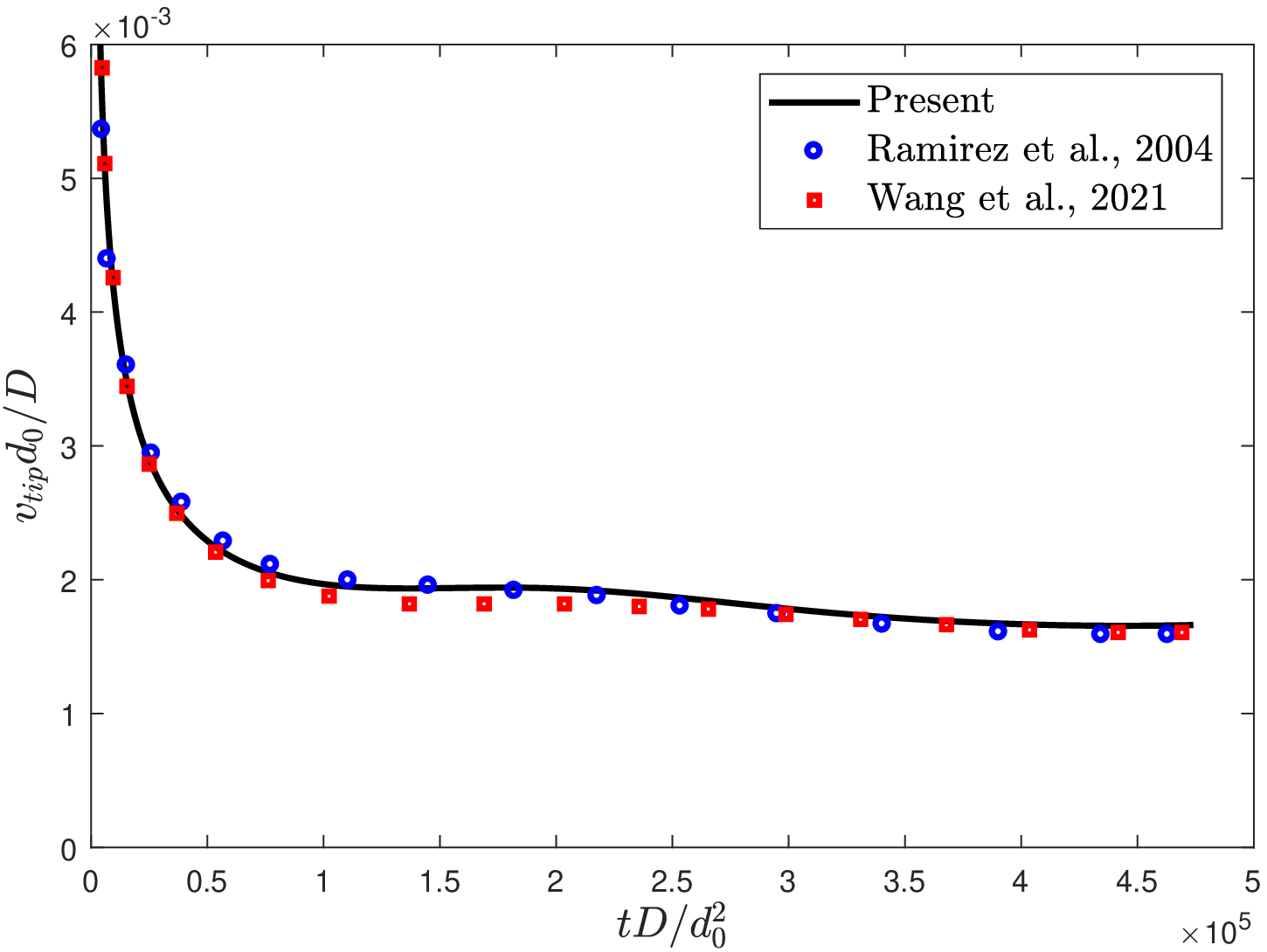}
			\label{fig-TCpure-Vtip}
	\end{minipage}}	
	\subfigure[]{
		\begin{minipage}{0.49\linewidth}
			\centering
			\includegraphics[width=3.0in]{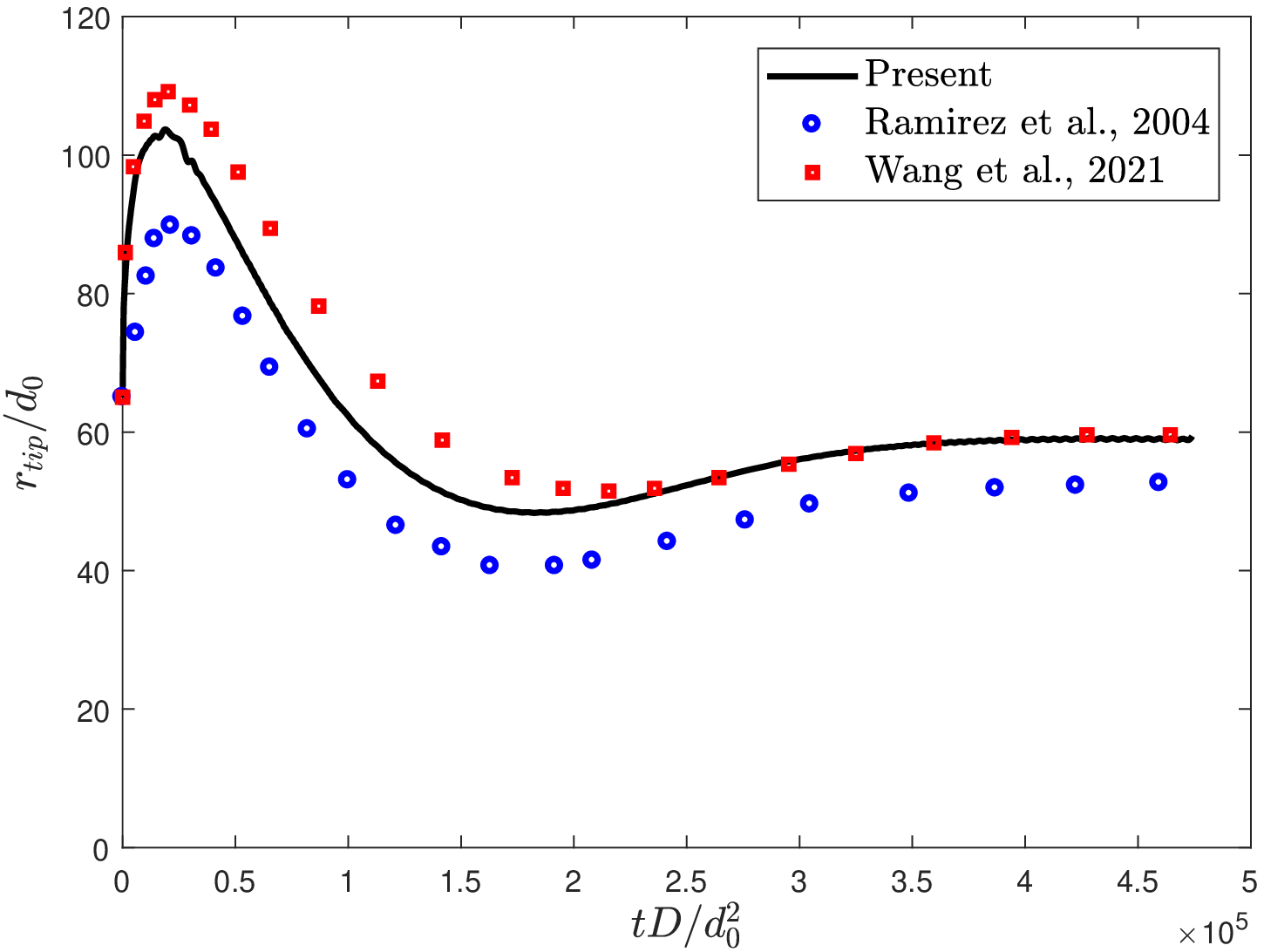}
			\label{fig-TCpure-Rtip}
	\end{minipage}}	
	\caption{Evolutions of tip velocity (a) and tip radius (b) in the thermosolutal dendritic growth with pure diffusion.}
	\label{fig-TCpure-tip}
\end{figure}
\begin{figure}
	\centering
	\subfigure[]{
		\begin{minipage}{0.49\linewidth}
			\centering
			\includegraphics[width=3.0in]{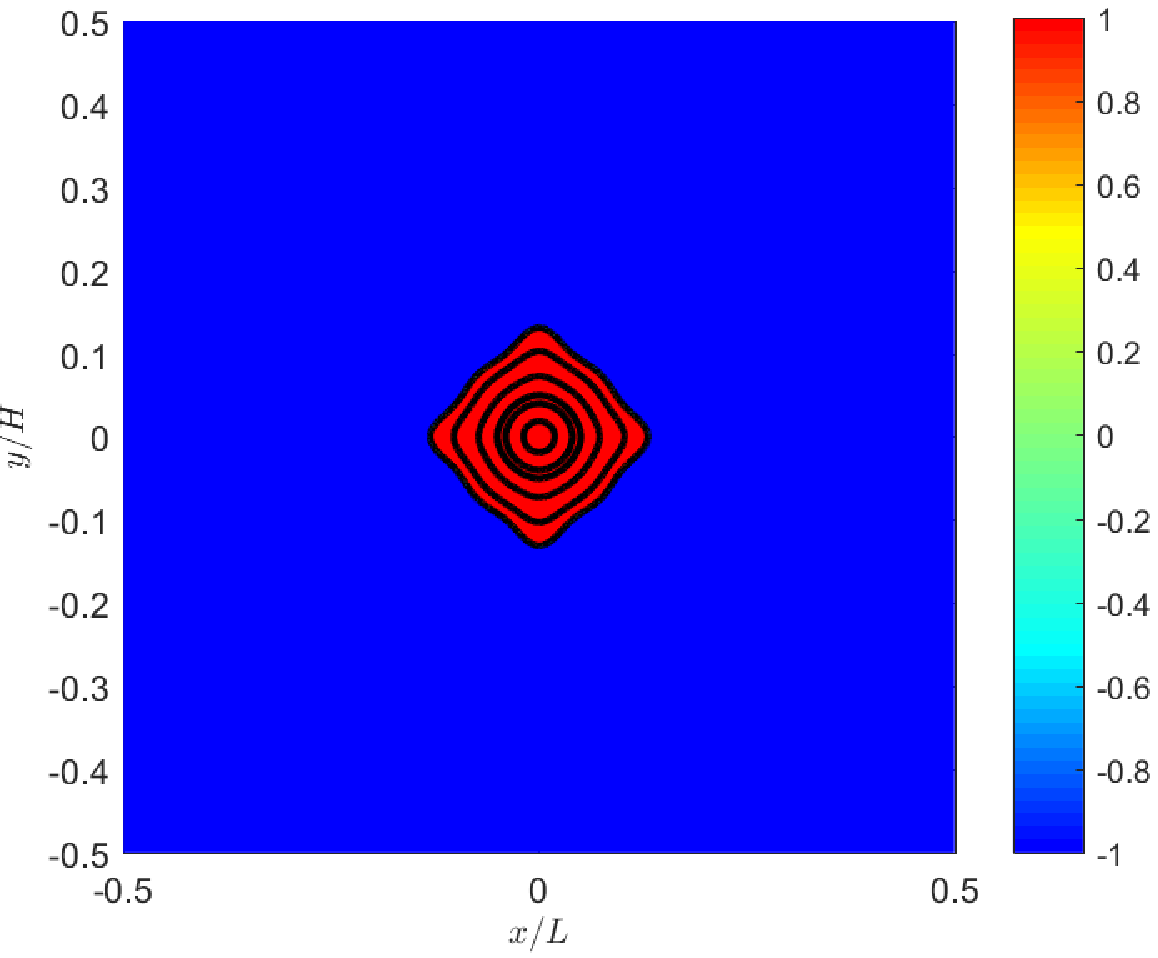}
			\label{fig-TCpure-phi}
	\end{minipage}}	
	\subfigure[]{
		\begin{minipage}{0.49\linewidth}
			\centering
			\includegraphics[width=3.0in]{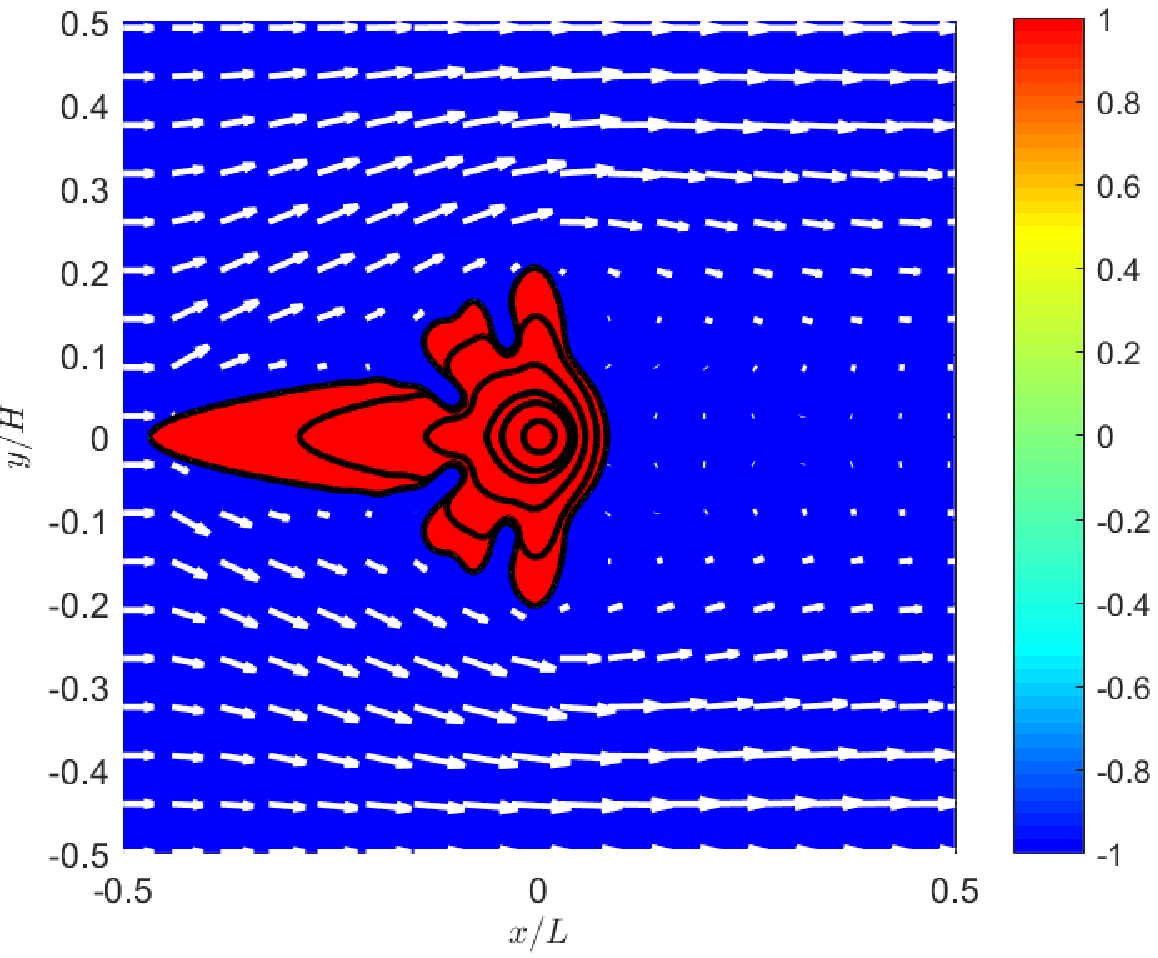}
			\label{fig-TCu-phi}
	\end{minipage}}	
	\subfigure[]{
		\begin{minipage}{0.49\linewidth}
			\centering
			\includegraphics[width=3.0in]{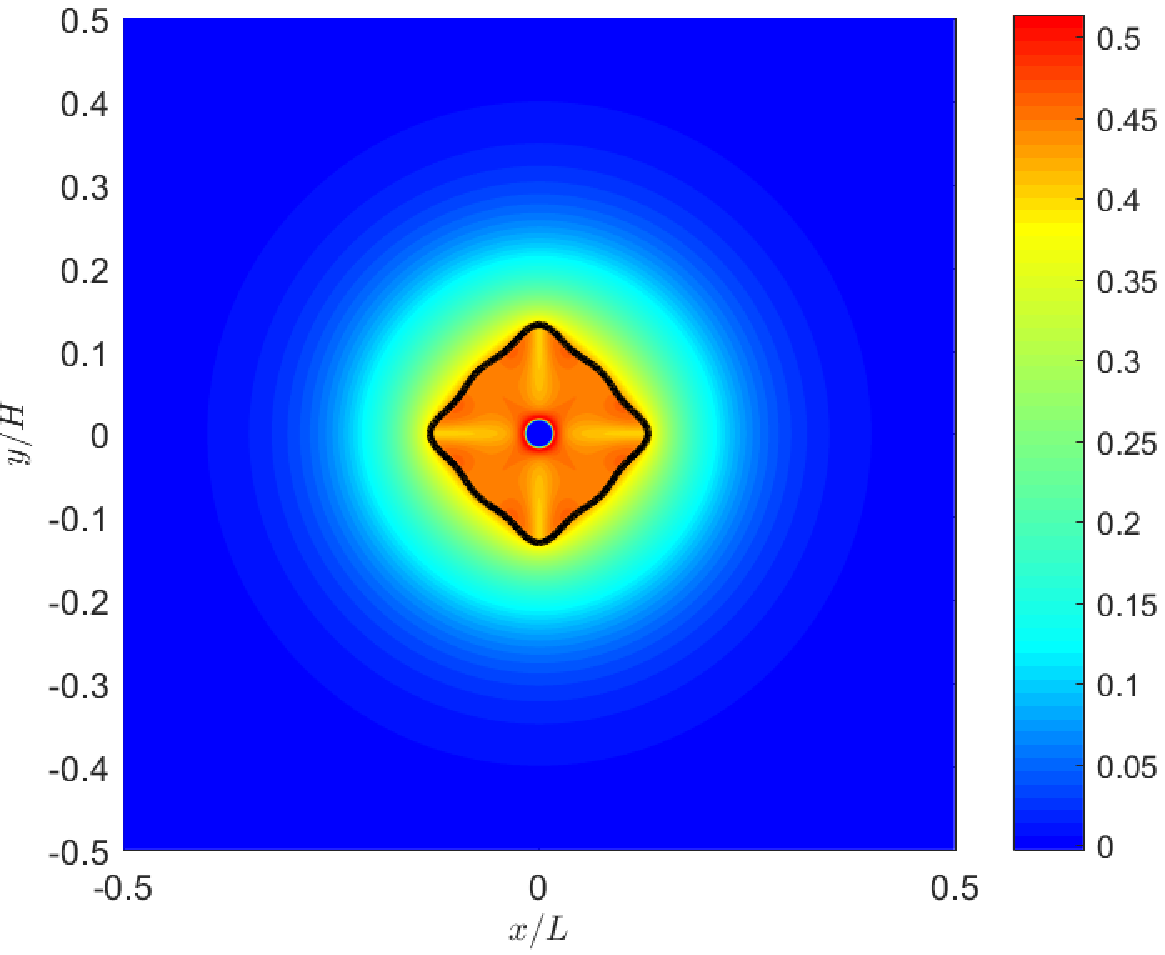}
			\label{fig-TCpure-C}
	\end{minipage}}	
	\subfigure[]{
		\begin{minipage}{0.49\linewidth}
			\centering
			\includegraphics[width=3.0in]{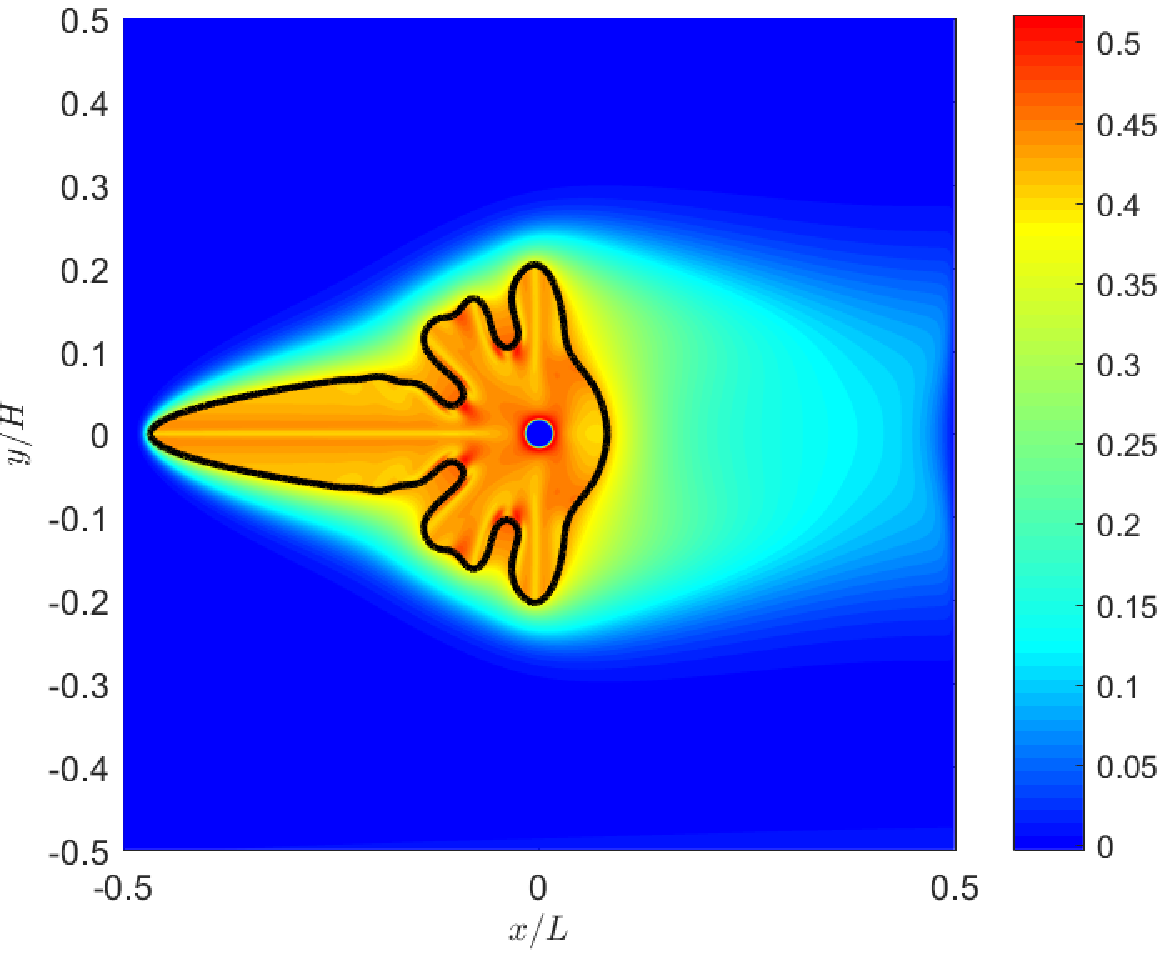}
			\label{fig-TCu-C}
	\end{minipage}}
	\subfigure[]{
		\begin{minipage}{0.49\linewidth}
			\centering
			\includegraphics[width=3.0in]{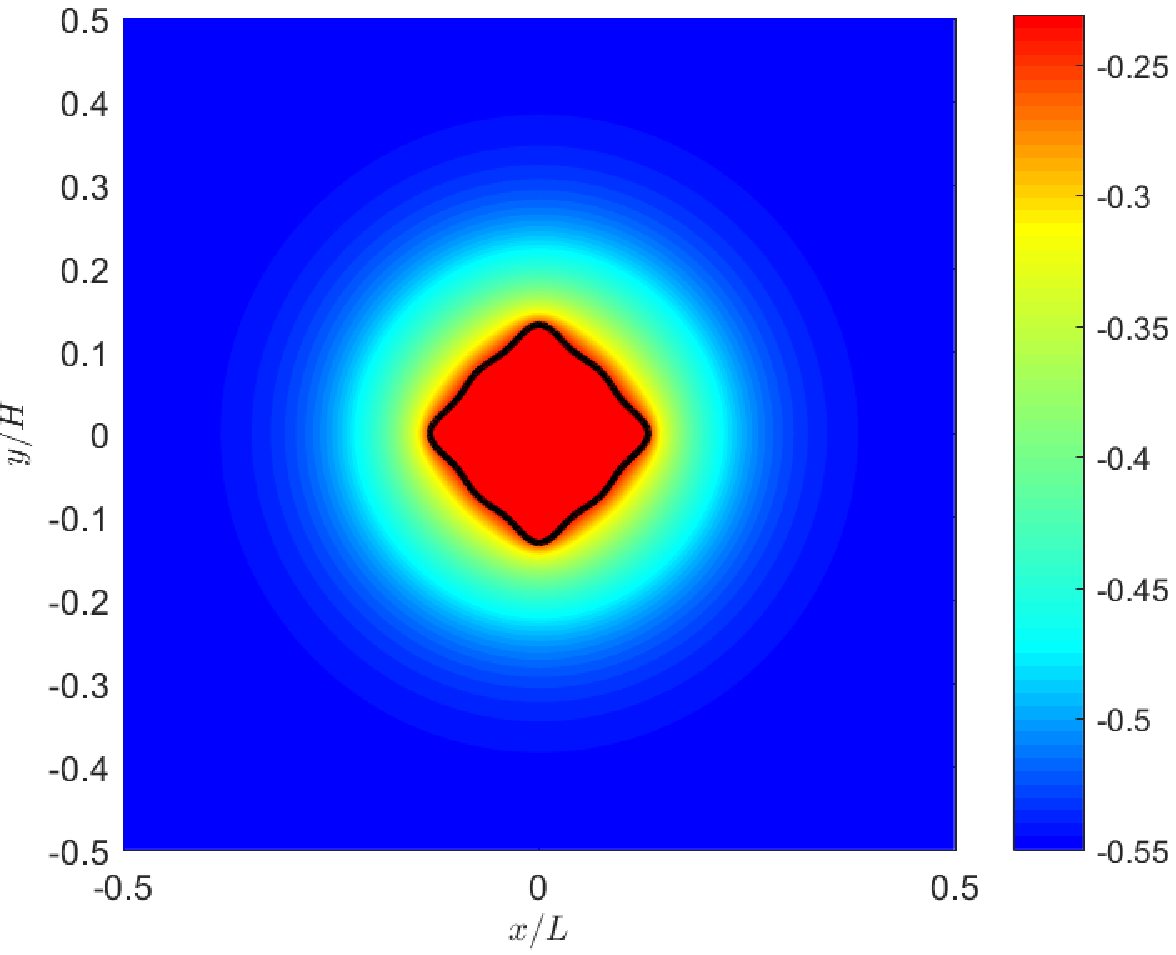}
			\label{fig-TCpure-T}
	\end{minipage}}	
	\subfigure[]{
		\begin{minipage}{0.49\linewidth}
			\centering
			\includegraphics[width=3.0in]{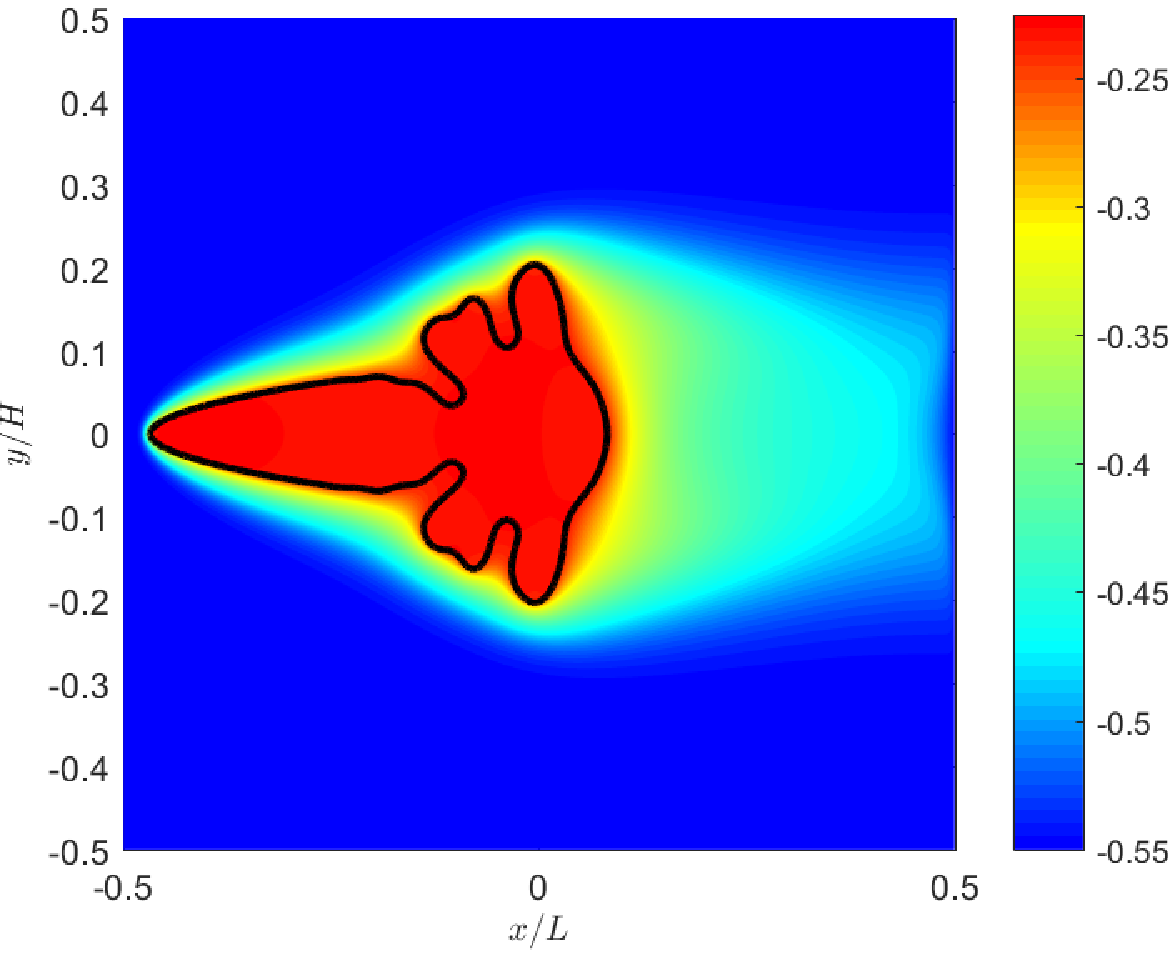}
			\label{fig-TCu-T}
	\end{minipage}}
	\caption{Snapshots of the phase field at $tD/d_0^2=0, 10000, 20000, 50000, 100000, 150000$ superimposed with velocity field at $tD/d_0^2=150000$ [(a): pure diffusion case, (b): melt flow case], the supersaturation field [(c): pure diffusion case, (d): melt flow case] and the temperature field at $tD/d_0^2=150000$ [(e): pure diffusion case, (f): melt flow case].}
	\label{fig-TC-com}
\end{figure}
Finally, the dendritic growth of a binary alloy in an undercooled melt is considered. For this problem, the heat and solute transfer as well as the melt flow work together, and more complex dendritic morphologies may be induced. 

For the thermosolutal dendritic growth with the pure diffusion, some parameters are taken as $L=2400\Delta x$, $\Delta t=\tau_0/55$, $\varepsilon_s=0.02$, $\theta_0=-0.55$, $U_0=0$, $\alpha=2$, $D=2$, $Mc_{\infty}=0.5325$, $k=0.15$, $\lambda=3.1913$ and $d_0=0.2770$. Here $Le=\alpha/D=1$ is considered for simplicity and other large/small values of $Le$ can also be simulated by present method with multi-time-scaling strategy. The initial seed radius is $R_s=45\Delta x$, and the far-field Dirichlet boundary conditions $\phi=-1$, $U=0$ and $\theta=-0.55$ are imposed. We first performed a comparison of the distributions of $\phi$, $U$ and $\theta$ along the central dendrite axis in Fig. \ref{fig-TCpure} where $tD/d_0^2=470000$. From this figure, one can observe that present results agree well with the reported data \cite{Ramirez2004PRE,Wang2021CMAME}. Then the evolutions of the tip velocity and radius are also shown in Fig. \ref{fig-TCpure-tip}, and it can be found that the tip velocity and radius are close to those in \cite{Ramirez2004PRE,Wang2021CMAME}. 

We continued to consider the thermosolutal dendritic growth coupled with melt flow, and the boundary conditions of flow field are the same as those stated previously. We presented the snapshots of phase field, supersaturation field and temperature field at $tD/d_0^2=150000$ in Fig. \ref{fig-TC-com}. As seen from this figure, the effect of melt flow on dendrite is more significant than those in subsections \ref{thermal} and \ref{solutal}. Actually, under the influence of the inlet velocity $\mathbf{u}_{in}$, the upstream arm grows very fast, and the vertical arms also have an obvious growth. On the contrary, the downstream arm is completely suppressed and the interface is much smoother. These results may be attributed to the large inlet velocity, which also brings the growth of the second dendrite arms. In addition, similar to the previous results in subsections \ref{thermal} and \ref{solutal}, the gradients of temperature and supersaturation around the upstream tip are much larger. 
 
\section{Conclusions}\label{Conclusion}
In this paper, a diffuse-interface LB method is proposed for the dendritic growth with thermosolutal convection. We first reformulated the governing equations for the solute and heat transfer by introducing a new variable related to the order parameter and temperature or supersaturation.
Then the diffuse-interface LB method is developed to treat the fluid-solid interaction such that the explicit interface tracking can be avoided. The present method is also tested by some classical problems, including the flow around a stationary circular cylinder, thermal dendritic growth, solutal dendritic growth, and thermosolutal dendritic growth with both pure diffusion and melt flow, and the numerical results are in good agreement with some available data. Finally, we would like to point out that the present method is also suitable for the problems of free moving dendritic growth, which would be considered in another work.    
\section*{Acknowledgments}
This research has been supported by the National Natural Science Foundation of China under Grants No. 12072127, No 51836003, No. 52075201 and No. 52188102.
\appendix
\section{Taylor expansion of the LB model for convection-diffusion type equation}\label{DTE}
In this appendix, the following general convection-diffusion equation is considered,
\begin{equation}\label{CDE}
	a_n\frac{\partial A}{\partial t}+\nabla\cdot\mathbf{B}=\nabla\cdot\left(\kappa\nabla C+\mathbf{J}\right)+S,
\end{equation}
where $A$ and $C$ are scalar variables, $\mathbf{B}$ is the convection term, $\kappa$ is the diffusion coefficient, $\mathbf{J}$ represents a coupled flux term, and $S$ denotes the source term.

Applying the Taylor expansion to Eq.\,(\ref{LBE}), we have \cite{Chai2020PRE}
\begin{equation}
	\sum_{l=1}^{N}\frac{\Delta t^l}{l!}\tilde{D}_i^lf_i+O\left(\Delta t^{N+1}\right)=-\Lambda_{ij}f_j^{ne}+\Delta t\left(F_i+\frac{\Delta t}{2}\hat{D}_iF_i\right)+\Delta t\left(\delta_{ij}-\frac{\Lambda_{ij}}{2}\right)G_j\left(\mathbf{x},t\right),
\end{equation}
where $\tilde{D}=a_n\partial_t+\mathbf{c}_i\cdot\nabla$ and $f_i^{ne}=f_i-f_i^{eq}$. From the above equation, one can obtain
\begin{subequations}
	\begin{equation}
		f_i^{ne}=O\left(\Delta t\right),
	\end{equation}
	\begin{equation}
		\sum_{l=1}^{N-1}\frac{\Delta t^l}{l!}\tilde{D}_i^l\left(f_i^{eq}+f_i^{ne}\right)+\frac{\Delta t^N}{N!}\tilde{D}_i^Nf_i^{eq}=-\Lambda_{ij}f_j^{eq}+\Delta t\left(F_i+\frac{\Delta t}{2}\hat{D}_iF_i\right)+\Delta t\left(\delta_{ij}-\frac{\Lambda_{ij}}{2}\right)G_j\left(\mathbf{x},t\right)+O\left(\Delta t^{N+1}\right),
	\end{equation}
\end{subequations}

Then, we can derive equations at different orders of $\Delta t$,
\begin{subequations}
	\begin{equation}\label{Odt}
		\tilde{D}_if_i^{eq}=-\frac{\Lambda_{ij}}{\Delta t}f_j^{ne}+F_i+\left(\delta_{ij}-\frac{\Lambda_{ij}}{2}\right)G_j+O\left(\Delta t\right),
	\end{equation}
	\begin{equation}\label{Odt2}
		\tilde{D}_i\left(f_i^{eq}+f_i^{ne}\right)+\frac{\Delta t}{2}\tilde{D}_i^2f_i^{eq}=-\frac{\Lambda_{ij}}{\Delta t}f_j^{ne}+F_i+\frac{\Delta t}{2}\hat{D}_iF_i+\left(\delta_{ij}-\frac{\Lambda_{ij}}{2}\right)G_j+O\left(\Delta t^2\right).
	\end{equation}
\end{subequations}
From Eq.\,(\ref{Odt}), we can get
\begin{equation}\label{DiOdt}
	\frac{\Delta t}{2}\tilde{D}_i^2f_i^{eq}=-\frac{1}{2}\tilde{D}_i\Lambda_{ij}f_j^{ne}+\frac{\Delta t}{2}\tilde{D}_i\left[F_i+\left(\delta_{ij}-\frac{\Lambda_{ij}}{2}\right)G_j\right]+O\left(\Delta t^2\right).
\end{equation}
Substituting Eq.\,(\ref{DiOdt}) into Eq.\,(\ref{Odt2}) yields
\begin{equation}\label{Odt2new}
	\tilde{D}_if_i^{eq}+\tilde{D}_i\left(\delta_{ij}-\frac{\Lambda_{ij}}{2}\right)f_j^{ne}+\frac{\Delta t}{2}\tilde{D}_i\left[F_i+\left(\delta_{ij}-\frac{\Lambda_{ij}}{2}\right)G_j\right]=-\frac{\Lambda_{ij}}{\Delta t}f_j^{ne}+F_i+\frac{\Delta t}{2}\hat{D}_iF_i+\left(\delta_{ij}-\frac{\Lambda_{ij}}{2}\right)G_j+O\left(\Delta t^2\right).
\end{equation}
To recover the correct governing equation (\ref{CDE}), the collision matrix and distribution functions should satisfy the following conditions,
\begin{subequations}
	\begin{equation}\label{sumLambda}
		\sum_ie_i\Lambda_{ij}=s_0e_j,\quad \sum_i\mathbf{c}_i\Lambda_{ij}=s_{10}e_j+s_1\mathbf{c}_j,
	\end{equation}
	\begin{equation}
		\sum_if_i^{eq}=A,\quad \sum_i\mathbf{c}_if_i^{eq}=\mathbf{B},\quad \sum_i\mathbf{c}_i\mathbf{c}_if_i^{eq}=C\hat{c}_s^2\mathbf{I},
	\end{equation}
	\begin{equation}
		\sum_iF_i=S,\quad \sum_i\mathbf{c}_iF_i=\mathbf{0},
	\end{equation}
	\begin{equation}
		\sum_iG_i=0,\quad \sum_i\mathbf{c}_iG_i=a_n\partial_t\mathbf{B}-\hat{c}_s^2\mathbf{J}/\kappa,
	\end{equation}
\end{subequations}
where $e_i=1$ for all $i=0,1,\cdots,q-1$, $s_0$, $s_{10}$ and $s_1$ are the eigenvalues of $\bm{\Lambda}$ for the eigenvectors $\left(e_i\right)$ and $\left(\mathbf{c}_i\right)$ \cite{Chai2020PRE}, respectively.

Summing Eq.\,(\ref{Odt}) and Eq.\,(\ref{Odt2new}) over $i$, we can obtain
\begin{subequations}
	\begin{equation}
		a_n\partial_tA+\nabla\cdot\mathbf{B}=S+O\left(\Delta t\right),
	\end{equation}
	\begin{equation}\label{sumOdt2}
		a_n\partial_tA+\nabla\cdot\mathbf{B}+\nabla\cdot\left(1-\frac{s_1}{2}\right)\left[\sum_i\mathbf{c}_if_i^{ne}+\frac{\Delta t}{2}\left(a_n\partial_t\mathbf{B}-\hat{c}_s^2\frac{\mathbf{J}}{\kappa}\right)\right]=S+O\left(\Delta t^2\right),
	\end{equation}
\end{subequations}
where $\sum_i\mathbf{c}_if_i^{ne}$ can be obtained from Eq.\,(\ref{Odt}),
\begin{equation}\label{cifi1}
	\sum_i\mathbf{c}_if_i^{ne}=-s_1^{-1}\Delta t\left[a_n\partial_t\mathbf{B}+\hat{c}_s^2\nabla C-\left(1-\frac{s_1}{2}\right)\left(a_n\partial_t\mathbf{B}-\hat{c}_s^2\frac{\mathbf{J}}{\kappa}\right)\right]+O\left(\Delta t^2\right).
\end{equation}
Substituting Eq.\,(\ref{cifi1}) into Eq.\,(\ref{sumOdt2}), we can derive the convection-diffusion equation Eq.\,(\ref{CDE}) at the order of $O\left(\Delta t^2\right)$,
\begin{equation}
	a_n\partial_tA+\nabla\cdot\mathbf{B}=\nabla\cdot\left(\kappa\nabla C+\mathbf{J}\right)+S+O\left(\Delta t^2\right),
\end{equation}
where $\kappa=\left(s_1^{-1}-1/2\right)\hat{c}_s^2\Delta t$.

Additionally, from Eq. (\ref{cifi1}), one can also obtain the local computing scheme for the gradient of $C$,
\begin{equation}\label{localDphi}
	\nabla C=-\frac{s_1}{\hat{c}_s^2\Delta t}\left(\sum_i\mathbf{c}_if_i-\mathbf{B}+\mathbf{J}+\frac{\Delta t}{2}a_n\partial_t\mathbf{B}\right).
\end{equation}

It should be noted that through choosing the specific meanings of the variables $a_n$, $A$, $\mathbf{B}$, $C$, $\kappa$ and $S$, one can get different convection-diffusion equations.
 
\section{The transformation matrix and moments of the used lattice models}\label{D2Q59}
In this MRT-LB model, the collision matrix is related to the transformation matrix $\mathbf{M}$, and moments of all used lattice models are listed below.

\noindent D2Q5 lattice model:

\begin{equation}
	\mathbf{M}=\begin{pmatrix}
		1 &  1 &  1 &  1 &  1 \\
		0 &  1 &  0 & -1 &  0 \\
		0 &  0 &  1 &  0 & -1 \\
		0 &  1 &  0 &  1 &  0 \\
		0 &  0 &  1 &  0 &  1 \\		
	\end{pmatrix},\quad
	\mathbf{\Lambda}=\mathbf{M}^{-1}\mathbf{SM}=\begin{pmatrix}
		s_0 &  s_0-s_2 &  s_0-s_2 &  s_0-s_2 &  s_0-s_2 \\
		0 &  \frac{s_1+s_2}{2} &  0 & \frac{-s_1+s_2}{2} &  0 \\
		0 &  0 &  \frac{s_1+s_2}{2} &  0 & \frac{-s_1+s_2}{2} \\
		0 &  \frac{-s_1+s_2}{2} &  0 &  \frac{s_1+s_2}{2} &  0 \\
		0 &  0 &  \frac{-s_1+s_2}{2} &  0 &  \frac{s_1+s_2}{2} \\		
	\end{pmatrix},
\end{equation}
where the diagonal relaxation matrix is given by $\mathbf{S}=\mathbf{diag}\left(s_0,s_1,s_1,s_2,s_2\right)$ and lattice sound speed $\hat{c}_s=\hat{c}/\sqrt{3}$ with $\hat{c}=\Delta x/\Delta t$.

\noindent D2Q9 lattice model:

\begin{equation}
\mathbf{M}=\begin{pmatrix}
	1 &  1 &  1 &  1 &  1 &  1 &  1 &  1 &  1\\
	0 &  1 &  0 & -1 &  0 &  1 & -1 & -1 &  1\\
	0 &  0 &  1 &  0 & -1 &  1 &  1 & -1 & -1\\
	0 &  1 &  0 &  1 &  0 &  1 &  1 &  1 &  1\\
	0 &  0 &  1 &  0 &  1 &  1 &  1 &  1 &  1\\	
	0 &  0 &  0 &  0 &  0 &  1 & -1 &  1 & -1\\			
	0 &  0 &  0 &  0 &  0 &  1 &  1 & -1 & -1\\
	0 &  0 &  0 &  0 &  0 &  1 & -1 & -1 &  1\\
	0 &  0 &  0 &  0 &  0 &  1 &  1 &  1 &  1\\
\end{pmatrix}.
\end{equation}
The diagonal relaxation matrix is $\mathbf{S}^{\mathbf{u}}=\mathbf{diag}\left(s_0^\mathbf{u},s_1^\mathbf{u},s_1^\mathbf{u},s_2^\mathbf{u},s_2^\mathbf{u},s_2^\mathbf{u},s_3^\mathbf{u},s_3^\mathbf{u},s_4^\mathbf{u}\right)$, $K=2\left(s_2^{\mathbf{u}}-3s_4^{\mathbf{u}}+s_2^{\mathbf{u}}s_4^{\mathbf{u}}\right)/9s_2^{\mathbf{u}}s_4^{\mathbf{u}}$.  

For the flow field, we can obtain the following moments,
\begin{equation}
	\begin{aligned}
		&\mathbf{m}_h^{eq}=\left[\rho_0,
		\frac{u}{\hat{c}},
		\frac{v}{\hat{c}},
		\frac{p+u^2}{\hat{c}^2},
		\frac{p+v^2}{\hat{c}^2},
		\frac{uv}{\hat{c}^2},
		\frac{v}{3\hat{c}},
		\frac{u}{3\hat{c}},
		\frac{p+u^2+v^2}{3\hat{c}^2}\right]^T,\\
		&\mathbf{m}_F=\left[0,
		\frac{F_x+f_x}{\hat{c}},
		\frac{F_y+f_y}{\hat{c}},
		2\varphi\frac{F_xu}{\hat{c}^2},
		2\varphi\frac{F_yv}{\hat{c}^2},
		\varphi\frac{F_xv+F_yu}{\hat{c}^2},
		\frac{F_y+f_y}{3\hat{c}},
		\frac{F_x+f_x}{3\hat{c}},
		2\varphi\frac{F_xu+F_yv}{3\hat{c}^2}\right]^T.
	\end{aligned}
\end{equation}
\bibliographystyle{elsarticle-num} 
\bibliography{ref}
\end{document}